\def\year{2022}\relax
\documentclass[letterpaper]{article} 
\usepackage{aaai22}  
\usepackage{times}  
\usepackage{helvet}  
\usepackage{courier}  
\usepackage[hyphens]{url}  
\usepackage{graphicx} 
\def\UrlFont{\rm}  
\usepackage{natbib}  
\usepackage{caption} 
\DeclareCaptionStyle{ruled}{labelfont=normalfont,labelsep=colon,strut=off} 
\usepackage{amsmath}
\usepackage{algorithm}
\usepackage{algorithmic}
\usepackage{pgf}
\usepackage{tikz}
\usepackage{tikzscale}
\usepackage{pgfplots}
\usepackage{caption}
\usepackage{subcaption}
\usepackage{wrapfig}

\usepackage{import}
\usepackage{xcolor}

\usetikzlibrary{quotes,arrows.meta}
\usetikzlibrary{positioning}

\def\edgecolor{rgb:blue,4;red,1;green,4;black,3}
\newcommand{\midarrow}{\tikz \draw[-Stealth,line width =0.8mm,draw=\edgecolor] (-0.3,0) -- ++(0.3,0);}

\usepackage{Ball}
\usepackage{Box}
\usepackage{RightBandedBox}

\usetikzlibrary{positioning}
\usetikzlibrary{3d} 

\def\InputColor{rgb:blue,5;green,1;white,3}
\def\ConvColor{rgb:yellow,5;red,2.5;white,5}

\def\PoolColor{rgb:red,1;black,0.3}

\def\FcColor{rgb:blue,5;red,2.5;white,5}
\def\FcReluColor{rgb:blue,5;red,5;white,4}
\def\SoftmaxColor{rgb:magenta,5;black,7}
\def\ConcatColor{rgb:blue,5;red,2.5;white,5}

\usepackage{newfloat}
\usepackage{listings}
\floatstyle{ruled}
\newfloat{listing}{tb}{lst}{}
\floatname{listing}{Listing}
\nocopyright
\title{Computing an Optimal Pitching Strategy in a Baseball At-Bat}
\author{
Connor Douglas,\textsuperscript{\rm 1}
Everett Witt,\textsuperscript{\rm 2}
Mia Bendy,\textsuperscript{\rm 3}
Yevgeniy Vorobeychik\textsuperscript{\rm 1}\\
\textsuperscript{\rm 1}Washington University in St. Louis,
\textsuperscript{\rm 2}Periwinkle Trading LLC, 
\textsuperscript{\rm 3}Capital One\\
\{c.douglas,everett.witt,mia.bendy,yvorobeychik\}@wustl.edu
}

\begin{document}

\maketitle

\begin{abstract}
The field of quantitative analytics has transformed the world of sports over the last decade. To date, these analytic approaches are statistical at their core, characterizing what is and what was, while using this information to drive decisions about what to do in the future. However, as we often view team sports, such as soccer, hockey, and baseball, as pairwise win-lose encounters, it seems natural to model these as zero-sum games. We propose such a model for one important class of sports encounters: a baseball at-bat, which is a matchup between a pitcher and a batter. Specifically, we propose a novel model of this encounter as a zero-sum stochastic game, in which the goal of the batter is to get on base, an outcome the pitcher aims to prevent. The value of this game is the on-base percentage (i.e., the probability that the batter gets on base). In principle, this stochastic game can be solved using classical approaches. The main technical challenges lie in predicting the distribution of pitch locations as a function of pitcher intention, predicting the distribution of outcomes if the batter decides to swing at a pitch, and characterizing the level of patience of a particular batter. We address these challenges by proposing novel pitcher and batter representations as well as a novel deep neural network architecture for outcome prediction. Our experiments using Kaggle data from the 2015 to 2018 Major League Baseball seasons demonstrate the efficacy of the proposed approach.
\end{abstract}

\section{Introduction}
Baseball is one of the most popular team sports in the U.S., with Major League Baseball (MLB) bringing in over \$10 billion in revenue in 2019~\citep{Bbrev20}.
Moreover, baseball has considerable global popularity as well, particularly in Latin America, Japan, and South Korea. 
With such a large market, teams look to gain a competitive edge, and to this end
they leverage complex statistical models generated from increasingly abundant baseball data. 
Yet, the scope of baseball analytics has been limited primarily to statistical and machine learning approaches, rather than game-theoretic reasoning~\citep{Alcorn17,Koseler17}.
There have been several proposals for strategic analysis of sports competitions outside of baseball, such as tennis and soccer.
\citet{Walker01} study a game-theoretic model of server-receiver interaction in tennis, but focus largely on indirectly evaluating the extent to which players act according to a mixed-strategy Nash equilibrium.
\citet{Azar11} consider a simple one-shot game-theoretic model of a soccer penalty kick, and several recent efforts consider a game-theoretic model of both tactical and strategic decisions in an actual soccer match~\citep{Beal20,Beal21}.
The latter, however, side-step complete strategic reasoning by estimating opponent strategies from historical data and then optimizing the strategy for a target team.


We present the first (to our knowledge) game-theoretic model of strategic interactions in a baseball game.
Our focus is a baseball at-bat, an encounter between a pitcher and a batter.
In an at-bat, a pitcher throws a series of pitches to a batter.
Every at-bat ends in one of two ways: 1) the batter is out (and, in our model, the pitcher wins), for example, after receiving the third strike, or 2) the batter gets on-base, for example, by hitting a home run.
In modeling an at-bat, we assume that the goal of the batter is solely to get on base, while the pitcher aims to get the batter out.
This focus on the on-base-percentage (OBP) is clearly restrictive (in not accounting, say, for the difference between walks and home runs), but is nevertheless an important element of baseball analytics~\citep{Albert02,Lewis04}.

We model an at-bat as a stochastic game in which the count (of balls and strikes) serves as the state, the pitcher's actions amount to which pitch to throw and where, while the batter decides whether to swing or take (not swing at) the pitch.
This model introduces two principal conceptual and technical challenges.
The first is that the pitcher and batter decisions are not, in fact, concurrent: the batter does get to observe the pitch as it leaves the pitcher's hand and as it travels towards home plate.
On the other hand, the relevant observation window is so short (usually less than half a second) that the batter has little ability to deliberate upon their decision.
The particularly salient issue here is that no batter will swing if the pitch is far outside of the strike zone, while when pitches are close, batters vary significantly in their swing propensity (what is often called a batter's ``patience'' or ``eye'').
The second challenge is that our model governs the pitcher's intent about where the pitch is thrown.
However, pitching data documents only where the pitches \emph{ended up}; we observe nothing explicit about intent.

One way to deal with the challenge of observability is to move to a full-blown partially-observable stochastic game (POSG) model.
However, POSGs are notoriously difficult to solve.
We propose instead to add an explicit element to our model that allows us to both capture the most salient nature of this partial observability while keeping the main stochastic game structure.
Specifically, we use data to learn a batter-specific probability that a batter swings at (relatively) borderline pitches outside the strike zone, and when this is sufficiently high, ``override'' a batter's decision to swing prescribed by the model by modifying the associated transition to always result in a ball.

We address the second challenge by first decomposing the distribution of outcomes given the pitcher's actions, conditional on the batter swinging, into two parts: 1) distribution of \emph{actual} locations given \emph{intended} locations, and 2) distribution of outcomes based on actual locations.
We learn the former by assuming that the error distribution of the pitcher is Gaussian and by taking advantage of counts in which most pitchers aim to throw a strike most of the time.
For the latter, we propose a novel tensor representation of the pitcher and batter, along with a novel deep neural network architecture, and use historical baseball data to learn the outcome distribution for given pitcher-batter pairs.
With all the pieces in place, we can  solve the resulting stochastic game using standard methods~\citep{Filar12,Littman94}.

We evaluate the parts of our approach, as well as the final game theoretic equilibrium pitch distribution, using 2015-2018 Kaggle data for Major League Baseball.
We show that our deep neural network models can successfully capture distributions of outcomes as well as batters' patience, and demonstrate that the proposed approach yields significantly lower predicted OBP for the batters (i.e., higher utility for the pitcher) than the empirical OBP in most counts.

\smallskip
\noindent{\bf Related Work }
With team sports typically being win-lose encounters between a pair of teams, it is natural to model these using the language of game theory.
Indeed, some individual sports, such as tennis, have a similar two-player win-lose (zero-sum) element.
Consequently, several approaches to game-theoretic modeling of sports activities have been previously proposed.
For tennis specifically, \citet{Walker01} studied empirically whether top players at Wimbledon play each point according to a mixed-strategy equilibrium.
They showed that on the one hand, evidence appears to support this in that the probability of winning a point is similar whether the server serves to the right or the left of the receiver.
On the other hand, aspects such as that server's decisions whether to serve to the left or to the right, do not appear to be independent random draws as the game-theoretic model would predict.
In a more theoretical effort, \citet{Walker11} propose a stochastic game model of win-lose encounters in which the pivotal quantity is a score, which determines the nature of each stage game, and stage games have only two possible outcomes determined by which of the two players win.
This stochastic game model is related to ours, but our model has four possible outcomes in each state, violating one central assumptions of \citet{Walker11}; more fundamentally, our model of a baseball at-bat does not fit their general assumed structure of the stochastic game as transitioning through a series of ``point games''.
Moreover, our goals are different: while \citet{Walker11} focus on characterizing the structure of equilibria in this game, our goal is to solve it, and infer the structure of the game from data.

\citet{Azar11} study whether penalty kick interactions between the kicker and the goalie in soccer are representative of mixed-strategy Nash equilibrium play.
In this setting, however, there are no environment dynamics, and the game is straightforward to empirically represent.
Indeed, to the extent that outcome distribution depends on the particular players involved, \citet{Azar11} do not model this.
In contrast, this, along with the stochastic game model, is a central contribution of our work.
\citet{Beal20} and \citet{Beal21} study strategic and tactical decision making in a soccer game.
Their models are a blend of Bayesian and stochastic games, but the ultimate approaches do not investigate solutions to these games as such. Instead, they use data to estimate the strategy of a given opponent and then approach the problem as an optimization problem with several alternative objectives, rather than a game.

The increasing importance of sports analytics, particularly in baseball, has in turn given rise to a number of machine learning approaches surveyed by~\citet{Koseler17}.
However, the classes of problems investigated in such approaches have been so far typically limited to predicting which pitch will be thrown~\citep{Ganeshapillai12,Hoang15}, a player's batting average or other offensive statistics~\citep{Jiang10,Lyle07}, likelihood of catching a baseball~\citep{Das94}, likelihood of winning~\citep{Yang04}, and the like.
There do not appear to be prior approaches to predict the outcomes for a pair of pitcher and batter at the level of resolution of a single pitch, which is one of our main contributions.
Finally, \citet{Alcorn17} proposed an approach for real vector embedding of baseball players in the style of word2vec~\citep{Mikolov13}.
We also learn player embeddings, with the important difference that the embedding by \citet{Alcorn17} assumes that the set of players is \emph{static}, an assumption routinely violated in MLB; in contrast, we make no such assumption.
\citet{Sarris18} discuss approaches for characterizing pitcher control, but to date the focus is predominantly on simple aggregate statistics, such as median error, rather than inferring the full distribution as we do (and which is crucial to our overall approach).

Finally, more loosely related are the efforts to use computational game theory to solve complex canonical games, such as poker~\citep{Brown18,Brown19}.
However, aside from the obvious difference in the game structure between poker and baseball, conceptually the key distinction is that in poker and similar games, the game itself, including player actions and payoffs, is entirely specified and common knowledge.
In contrast, a game-theoretic model of baseball (or a baseball at-bat) as a game is not a priori self-evident and is in fact one of our central contributions.





\section{Background: the Basics of Baseball}


A (U.S. major league) baseball game is an encounter between two teams and proceeds through a series of nine \emph{innings}.
Each inning is comprised of two \emph{half-innings}: the \emph{top} half and the \emph{bottom} half.
In the top half-inning, the away team bats, while the home team pitches and defends, and the roles reverse in the bottom half-inning.
Each half-inning consists of three outs, that is, an inning proceeds until three players on the batting team have registered an out.

The most basic interaction in baseball is an at-bat, which is a faceoff between a pitcher, who throws a baseball towards the home plate area, and a batter standing near this area 
\begin{wrapfigure}{l}{0.25\textwidth}
    \centering
    \includegraphics[width=0.25\textwidth]{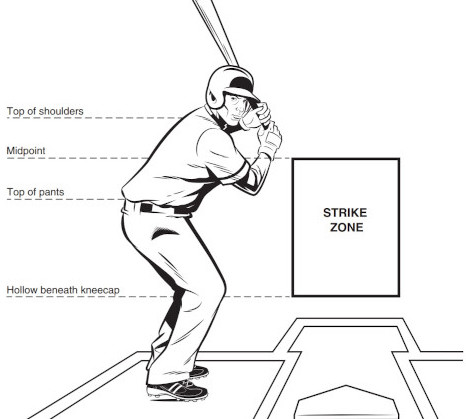}
    \caption{Visual illustration of a strike zone.}
    \label{F:sz}
\end{wrapfigure}
with the aim to ``get on base''.
A central concept in an
 at-bat is a \emph{strike zone} (see Figure~\ref{F:sz}).
The vertical range of the strike zone starts at just below the batter's knees and reaches up to the midpoint between the batter's shoulders and the top of his pants.
The horizontal range of the zone is the same as the width of the home plate---a white pentagon drawn on the ground with the flat bottom side facing the pitcher.

While the strike zone clearly depends on the batter, we'll treat it as a fixed entity for simplicity.
Any pitch that is in the strike zone when it crosses the home plate is considered a \emph{strike}, and when the batter observes, or swings through, strike 3, he is automatically out.
Conversely, a pitch that is outside the strike zone is called a \emph{ball}, and the batter is automatically on base whenever they observe a fourth ball.
A \emph{count} keeps track of the number of balls and strikes in an at-bat, starting at 0-0 (balls-strikes).

At any point, if a batter swings at a pitch, one of four things can happen: 1) a hit, which happens whenever the ball lands in the field of play, or is a home run (leaves the stadium within the field of play), and cannot be reached by a defensive fielder before the batter reaches first base; 2) an out, which happens whenever the a fielder either catches a hit ball on the fly, or can throw it to first base (i.e., to the defender standing with one foot on first base) prior to the batter stepping on it; 3) a strike, if the batter does not make contact with the baseball; and 4) a foul, if the batter makes contact, but the ball lands behind the field of play.

Our concern is a particular at-bat, and the goal is to determine the optimal pitching sequence, be it deterministic or stochastic in order to minimize the probability that the batter ends up on base---commonly known as the \emph{on-base percentage} or \emph{OBP} in baseball parlance.
\section{A Baseball At-Bat as a Stochastic Game}

Consider an \emph{at-bat}, an encounter between a pitcher and a batter.
We start with a 0-0 count (0 balls, 0 strikes).
The pitcher then throws a pitch, let's say, a 95-mph fastball.
As the fastball zips rapidly towards home plate, we hold our breath as we anticipate one of a number of yet-unrealized possibilities: perhaps the batter takes a strike, perhaps he swings and fouls the pitch off, or, perhaps, he makes hard contact, and the ball clears the fence in deep left field.
A home run, a base hit, or a walk, all end an at-bat, as does a strikeout.
However, a ball, a strike, and a foul ball may (in the latter case, will always) continue the at-bat, potentially changing the count.
For example, a strike or a foul ball in a 0-0 count will always progress the count to 0-1, and if a ball follows, the count progresses again to 1-1.
On a 3-2 count, a foul ball returns us to the same 3-2 count, while a ball necessarily results in a walk and a (non-foul) strike in an out.

Since there is no evident private information in an at-bat, it is natural to model it as a zero-sum stochastic game.
However, a good model is not obvious.
First, what are the action sets for the players?
For example, the batter does observe the pitch as it traverses the airspace between the pitcher's mound and the catcher's glove, and could use this information (e.g., spin, location, speed of the pitch) to decide whether, and how, to swing.
Aside from the considerable complexity this introduces, none of this information is available in public baseball data at the level of necessary detail.
Furthermore, what is state?
In principle, everything may matter: who is on base, how hard the wind is blowing, the stadium, the entire sequence of pitches that has been thrown in that at-bat, or in the game so far, and even what the batter had for breakfast.
Again, we clearly need an abstraction that allows us to reason about the interaction, as well as to inform the model based on available data.

We propose a novel model of a baseball at-bat as the following stochastic game involving two players, a pitcher and a batter.
First, we consider only two types of outcomes of at-bats: on-base (a hit or a walk) and out.
Thus, we do not differentiate between a single and a home run.

\begin{figure}[h!]
    \centering
    \includegraphics[width=1.2in]{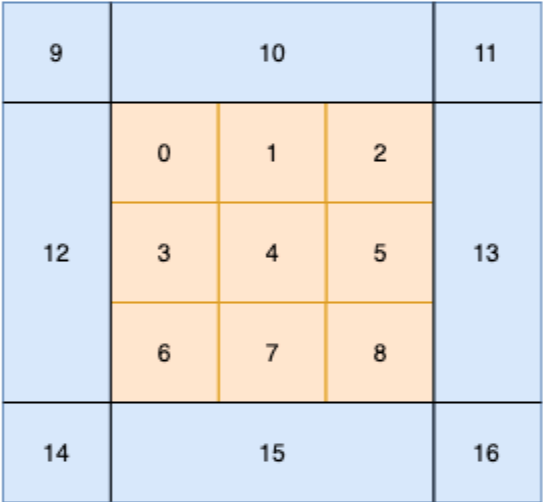}
    \caption{Pitch zone model from the pitcher's point of view: the orange center is the strike zone, while the blue periphery is outside the strike zone. The numbers are numerical labels for associated discrete zone segments.}
    \label{F:pz}
\end{figure}

Second, we define the actions of the two players as follows.
For the pitcher, let $P$ be the set of all pitch types they can throw, and $L$ the set of possible locations (both in the strike zone, and outside), which we assume to be finite (discretizing the strike zone, and the ``non-strike'' zone).
Let $Z_s \subset L$ be the strike zone, and $Z_b \subset L$ be locations outside the strike zone (see Figure~\ref{F:pz} for a particular discretization of both $Z_s$ and $Z_b$).
The pitcher's action space is then $A_p = P \times L$.
Note that in this context, locations $l \in L$ are \emph{intended} locations: that is, locations that the pitcher is \emph{aiming} for; of course, the pitch may often end up elsewhere, and we will come back to this below.
For the batter, we define actions simply as binary: whether to swing at the pitch (denoted by $\sigma$), or to take (not swing; $\tau$), that is, $A_b = \{\sigma,\tau\}$.
There is an important complication here: clearly, sometimes the pitch will be so far out of the zone that no batter would plausibly swing; we come back to this below as well.
We use $a = (a_p,a_b) \in A_p \times A_b$ to denote the action profile (joint actions) of both players.

Next, we define the state space $S$ as follows.
Let $u \in U \equiv \{0,1,2,3\}$ be the number of balls and $v \in V \equiv \{0,1,2\}$ the number of strikes, and let $b$ represent the state in which the batter is on-base, while $o$ the state with the batter being out.
The state space is then $S = U \times V \cup b \cup o$.
We denote states by $s \in S$.
Of course, other things may, and likely do, matter in an at-bat, but the information above is clearly first-order information, and therefore a natural place to start.

Since the game is zero-sum, it suffices to define the utility function $u(s,a)$, where $s \in S$ and $a \in A_p \times A_b$ just for the batter.
Moreover, since we only care about on-base or outs, $u(s,a) = 0$ for any $s \in U \times V$ (active at-bat), and $u(b,a) = 1$ while $u(o,a) = 0$.
Therefore, the expected utility in this game (i.e., probability the batter gets on base) has the natural interpretation as \emph{on-base percentage (OBP)}, which the goal of the batter is to maximize and the pitcher's goal is to minimize.

Our final task is to define the transition distribution $T_{ss'}^{a}$ in this stochastic game.
Central to this are two sources of uncertainty conditional on the action choices $a$ of both players: 1) uncertainty about where the chosen pitch $p \in P$ ends up given its intended location $l$, and 2) uncertainty about the outcome if the batter chooses to swing, i.e., $a_b = \sigma$.
Let's start with pitch location uncertainty.
Let $D_l(p,l)$ be the distribution over locations $l' \in L$ \emph{where the pitch ends up}, given that a pitch $p \in P$ was aimed at location $l \in L$.
To deal with outcome uncertainty, first note that if the batter takes, the outcome is fully determined by pitch location: either the pitch is in the strike zone, in which case it's a strike, or not, and it's a ball (we ignore here the additional complication of umpire mistakes).
If the batter swings, there are four possibilities: 1) the batter swings-and-misses (a strike), 2) foul ball, 3) hit, and 4) out after putting the ball in play.
Let $\Omega = \{\omega_s, \omega_f, \omega_h, \omega_o\}$ be the set of these four outcomes, respectively.
Let $D_\omega(p,l,s)$ be the probability distribution over $\Omega$ conditional on the batter swinging, if the pitch is $p$ and the \emph{actual} (not necessarily intended) location of the pitch is $l$ (it also in general depends on count; hence the explicit dependence on state $s$).

With $D_l$ and $D_\omega$ in hand, we can now completely define the transition distribution $T_{ss'}^{a}$.
If $a_b = \tau$, then transition is deterministic and only depends on the count and $D_l$: 1) if the actual location $l \in Z_s$, the pitch is a strike; then if $v = 2$, $s' = o$, and otherwise $v' = v+1$; 2) if $l \in Z_b$ (the pitch is a ball), then if $u = 3$, $s' = b$, and otherwise $u' = u+1$.
If the batter swings, transitions are now determined by $D_\omega$.
Thus, if the (stochastic) outcome $\omega = \omega_s$, the transition is exactly the same as if the batter took the strike.
If $\omega = \omega_f$, then if $v = 2$, $v' = 2$, and otherwise, $v' = v+1$.
If $\omega = \omega_h$, then $s' = b$ (if it's a hit, the batter ends up on base), and, finally, if $\omega = \omega_o$, then $s' = o$ (the batter is out).

If we are given all of the information above, including $D_l$ and $D_\omega$, we can solve this game using the combined value iteration and linear programming approach proposed by \citet{Littman94}, which we review below.
The challenge is that not only are these not given explicitly, but $D_l$ is pitcher-specific, and $D_\omega$ depends on the particular matchup between the pitcher and batter.

An additional challenge is the following: in our model above, we effectively assumed that the batter decides whether to swing or take \emph{at the same time} as the pitch is thrown.
In practice, batters can partially detect where the pitch will end up, a characteristic commonly referred to as a batter's \emph{patience} (their propensity to swing at pitches outside the strike zone).
Moreover, patience is certainly batter-specific: some batters are far more patient than others, and this is a crucial aspect to capture.
We model this feature by introducing a binary patience function $G(l)$ for $l \in Z_b$, where $G(l) = 0$ means that the batter will in fact take that pitch \emph{even if they intended to swing initially}, while $G(l) = 1$ means that the batter will proceed with swinging.
Of course, this, too, is not give a priori.

Next, we describe in detail how we can use baseball data to arrive at estimates of $D_l$, $D_\omega$, and $G$.
We then put everything together in solving the resulting stochastic game.
\section{Solution Approach}

We now describe our approach to computing an equilibrium of the stochastic game representing a baseball at-bat, given a specific pair of pitcher and batter.
We start by describing in detail how we learn the aspects of the game model that determine the state transition distribution from past data of baseball at-bats.
Subsequently, we describe how we solve the stochastic game using the well-known solution approach that combines linear programming with value iteration.

\subsection{Predicting Outcomes}
\label{S:predoutcomes}

The first missing piece of the game model that must be inferred from data is the distribution $D_\omega(p,l,s)$ that predicts outcomes for a pitch $p$ that is thrown (whether intended to or not) at location $l$ in state (count) $s$.
One of the central aspects of the game model is that this distribution also clearly depends on who is pitching, and who is batting.
Let $x_p$ and $x_b$ represent the pitcher and batter, respectively; our goal, more precisely, is to learn $D_\omega(p,l,s;x_p,x_b)$.

A naive idea is to represent the batter by either their batting average, on-base percentage, slugging percentage, or another of a myriad 1-dimensional attributes from baseball analytics, or even by concatenating several of these.
Similarly, one can represent a pitcher just using, say, their earned-run average.
However, such simplistic representations lose a great deal of information.
For example, it is often conventional to talk about some batters as, say, ``good low-ball hitters'', others as ``good fastball hitters'', and so on---clearly capturing information that is not simply reflected in simple aggregate statistics.
We therefore propose a novel representation of batters, and a novel representation of pitchers, aiming to capture as much readily available information about their past experience as possible, thereby avoiding making specific assumptions about them a priori.

\begin{figure}[h!]
    \centering
    \includegraphics[width=3in]{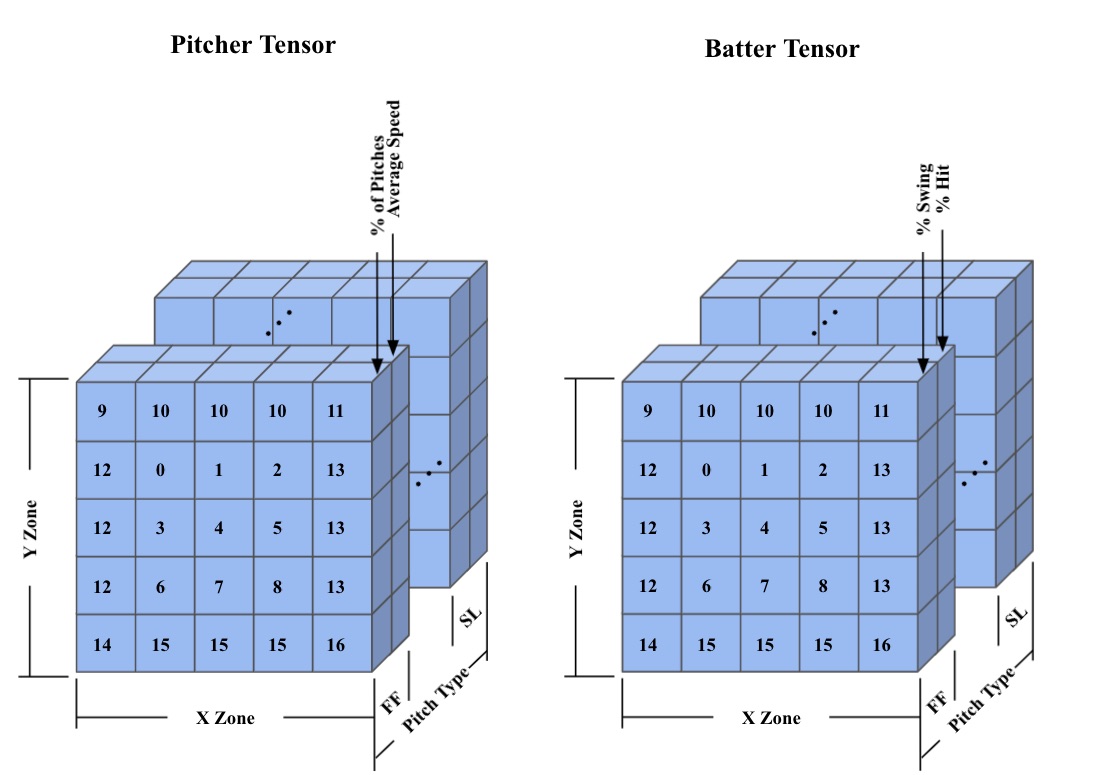}
    \caption{Pitcher and batter tensor visualizations. Both pitcher and batter tensors are of the shape (5,5,12) where the first two dimensions are used to physically represent the zone location. The final dimension is comprised of layers of statistics for each pitch type (pitch type thrown in the case of a pitcher and pitch type faced in the case of a batter).}
    \label{F:input_tensors}
\end{figure}
Let's start with batters.
We represent batters using a 3D tensor (Figure~\ref{F:input_tensors}) in which 2 dimensions correspond to the orientation of the (strike and ball) zone $Z$ (height and width; see Figure~\ref{F:pz}, but with zones 10, 12, 13, 15 split up into 3 each to correspond to a matrix, but with associated matrix entries taking on identical values) from the pitcher's perspective.
We then associate each pitch with a pair of slices in the third dimension: one for the relatively frequency of swinging at that pitch, and another for batting average if that pitch was thrown in the associated location.

Now, the pitchers.
We represent pitchers, just as batters (Figure~\ref{F:input_tensors}), using a 3D tensor with 2 dimensions corresponding to the zone.
We then associate each pitch with a pair of slices in the third dimension: one for the relative frequency of throwing that pitch in the particular location in the zone;
consequently, the sum of the entries over the zone and all pitches is 1; and the second for the average velocity of the associated pitch.

Input pitch type and location are represented by tensors with a similar shape, (5,5,6), to pitchers and batters, but use a one-hot encoding,
with a 1 in the position corresponding to the pitch and its location, and 0s elsewhere.
Finally, we represent count $s$ by two integers, one for the number of balls, and the other for the number of strikes.

\begin{figure}[h!]
\centering
\begin{tikzpicture}
\tikzstyle{connection}=[ultra thick,every node/.style={sloped,allow upside down},draw=\edgecolor,opacity=0.7]
\tikzstyle{copyconnection}=[ultra thick,every node/.style={sloped,allow upside down},draw={rgb:blue,4;red,1;green,1;black,3},opacity=0.7]



\pic[shift={(0,0,0)}] at (0,0,0) 
    {Box={
        name=inp0,
        caption=,
        ylabel={{pitcher}},
        fill=\InputColor,
        height=7,
        width=3,
        depth=7
        }
    };


\pic[shift={(0.5,0,0)}] at (inp0-east) 
    {Box={
        name=convp1a,
        caption= ,
        fill=\ConvColor,
        height=4,
        width=1,
        depth=4
        }
    };
    
\pic[shift={(0,0,0)}] at (convp1a-east) 
    {Box={
        name=convp1b,
        caption= ,
        fill=\ConvColor,
        height=4,
        width=1,
        depth=4
        }
    };
    
\pic[shift={(0,0,0)}] at (convp1b-east) 
    {Box={
        name=convp1c,
        caption= ,
        fill=\ConvColor,
        height=4,
        width=1,
        depth=4
        }
    };

\pic[shift={(0,0,0)}] at (convp1c-east) 
    {Box={
        name=poolp1,%
        caption= ,
        fill=\PoolColor,
        opacity=0.5,
        height=3,
        width=0.5,
        depth=3
        }
    };



\pic[shift={(0.5,0,0)}] at (poolp1-east) 
    {Box={
        name=convp2a,
        caption= ,
        fill=\ConvColor,
        height=4,
        width=1,
        depth=4
        }
    };
    
\pic[shift={(0,0,0)}] at (convp2a-east) 
    {Box={
        name=poolp2a,%
        caption= ,
        fill=\PoolColor,
        opacity=0.5,
        height=3,
        width=0.5,
        depth=3
        }
    };


\pic[shift={(0.1,0,0)}] at (poolp2a-east) 
    {Box={
        name=convp2b,
        caption= ,
        fill=\ConvColor,
        height=4,
        width=1,
        depth=4
        }
    };

\pic[shift={(0,0,0)}] at (convp2b-east) 
    {Box={
        name=poolp2b,%
        caption= ,
        fill=\PoolColor,
        opacity=0.5,
        height=3,
        width=0.5,
        depth=3
        }
    };
    

\pic[shift={(0.1,0,0)}] at (poolp2b-east) 
    {Box={
        name=convp2c,
        caption= ,
        fill=\ConvColor,
        height=4,
        width=1,
        depth=4
        }
    };

\pic[shift={(0,0,0)}] at (convp2c-east) 
    {Box={
        name=poolp2c,%
        caption= ,
        fill=\PoolColor,
        opacity=0.5,
        height=3,
        width=0.5,
        depth=3
        }
    };
    

\pic[shift={(0.1,0,0)}] at (poolp2c-east) 
    {Box={
        name=convp2d,
        caption= ,
        fill=\ConvColor,
        height=4,
        width=1,
        depth=4
        }
    };

\pic[shift={(0,0,0)}] at (convp2d-east) 
    {Box={
        name=poolp2d,%
        caption= ,
        fill=\PoolColor,
        opacity=0.5,
        height=3,
        width=0.5,
        depth=3
        }
    };
    

\pic[shift={(0.5,0,0)}] at (poolp2d-east) 
    {Box={
        name=Fcp1,
        caption= ,
        fill=\FcColor,
        height=5,
        width=1,
        depth=1
        }
    };
    
\draw [connection]  (inp0-east)    -- node {\midarrow} (convp1a-west);

\draw [connection]  (poolp1-east)    -- node {\midarrow} (convp2a-west);

\draw [connection]  (poolp2d-east)    -- node {\midarrow} (Fcp1-west);


\pic[shift={(0,-1,0)}] at (0,-1,0) 
    {Box={
        name=inb0,
        caption= ,
        ylabel={{batter}},
        fill=\InputColor,
        height=7,
        width=3,
        depth=7
        }
    };


\pic[shift={(0.5,0,0)}] at (inb0-east) 
    {Box={
        name=convb1a,
        caption= ,
        fill=\ConvColor,
        height=4,
        width=1,
        depth=4
        }
    };
    
\pic[shift={(0,0,0)}] at (convb1a-east) 
    {Box={
        name=convb1b,
        caption= ,
        fill=\ConvColor,
        height=4,
        width=1,
        depth=4
        }
    };
    
\pic[shift={(0,0,0)}] at (convb1b-east) 
    {Box={
        name=convb1c,
        caption= ,
        fill=\ConvColor,
        height=4,
        width=1,
        depth=4
        }
    };

\pic[shift={(0,0,0)}] at (convb1c-east) 
    {Box={
        name=poolb1,%
        caption= ,
        fill=\PoolColor,
        opacity=0.5,
        height=3,
        width=0.5,
        depth=3
        }
    };



\pic[shift={(0.5,0,0)}] at (poolb1-east) 
    {Box={
        name=convb2a,
        caption= ,
        fill=\ConvColor,
        height=4,
        width=1,
        depth=4
        }
    };
    
\pic[shift={(0,0,0)}] at (convb2a-east) 
    {Box={
        name=poolb2a,%
        caption= ,
        fill=\PoolColor,
        opacity=0.5,
        height=3,
        width=0.5,
        depth=3
        }
    };


\pic[shift={(0.1,0,0)}] at (poolb2a-east) 
    {Box={
        name=convb2b,
        caption= ,
        fill=\ConvColor,
        height=4,
        width=1,
        depth=4
        }
    };

\pic[shift={(0,0,0)}] at (convb2b-east) 
    {Box={
        name=poolb2b,%
        caption= ,
        fill=\PoolColor,
        opacity=0.5,
        height=3,
        width=0.5,
        depth=3
        }
    };
    

\pic[shift={(0.1,0,0)}] at (poolb2b-east) 
    {Box={
        name=convb2c,
        caption= ,
        fill=\ConvColor,
        height=4,
        width=1,
        depth=4
        }
    };

\pic[shift={(0,0,0)}] at (convb2c-east) 
    {Box={
        name=poolb2c,%
        caption= ,
        fill=\PoolColor,
        opacity=0.5,
        height=3,
        width=0.5,
        depth=3
        }
    };
    

\pic[shift={(0.1,0,0)}] at (poolb2c-east) 
    {Box={
        name=convb2d,
        caption= ,
        fill=\ConvColor,
        height=4,
        width=1,
        depth=4
        }
    };

\pic[shift={(0,0,0)}] at (convb2d-east) 
    {Box={
        name=poolb2d,%
        caption= ,
        fill=\PoolColor,
        opacity=0.5,
        height=3,
        width=0.5,
        depth=3
        }
    };
    

\pic[shift={(0.5,0,0)}] at (poolb2d-east) 
    {Box={
        name=Fcb1,
        caption= ,
        fill=\FcColor,
        height=5,
        width=1,
        depth=1
        }
    };
    
\draw [connection]  (inb0-east)    -- node {\midarrow} (convb1a-west);

\draw [connection]  (poolb1-east)    -- node {\midarrow} (convb2a-west);

\draw [connection]  (poolb2d-east)    -- node {\midarrow} (Fcb1-west);


\pic[shift={(0.5,0,0)}] at (Fcb1-east) 
    {Ball={
        name=cat1,
        caption=,
        fill=\ConcatColor,
        radius=1,
        logo=$||$
        }
    };



\pic[shift={(1,-2,0)}] at (0.5,-2,0)
    {Box={
        name=inpitch,
        caption= ,
        ylabel={{pitch}},
        fill=\InputColor,
        height=7,
        width=3,
        depth=7
        }
    };
    

\pic[shift={(1,0,0)}] at (inpitch-east) 
    {Box={
        name=convpitch,
        caption= ,
        fill=\ConvColor,
        height=4,
        width=1,
        depth=4
        }
    };

\pic[shift={(0,0,0)}] at (convpitch-east) 
    {Box={
        name=poolpitch,%
        caption= ,
        fill=\PoolColor,
        opacity=0.5,
        height=3,
        width=0.5,
        depth=3
        }
    };
    
\pic[shift={(0.95,0,0)}] at (poolpitch-anchor) 
    {Box={
        name=Fcpitch,
        caption= ,
        fill=\FcColor,
        height=5,
        width=1,
        depth=1
        }
    };
    
\draw [connection]  (inpitch-east)    -- node {\midarrow} (convpitch-west);

\draw [connection]  (poolpitch-east)    -- node {\midarrow} (Fcpitch-west);
    
    
\pic[shift={(-0.1,-1,0)}] at (Fcpitch-anchor) 
    {Box={
        name=Fccount,
        caption=,
        ylabel={{count}},
        fill=\FcColor,
        height=1,
        width=1,
        depth=1
        }
    };


\draw [connection]  (Fcb1-east)    -- node {\midarrow} (cat1-west);

\draw [connection]  (Fcp1-east) -- node {\midarrow} (Fcp1-east -| cat1-north) -- node {\midarrow} (cat1-north);

\draw [connection]  (Fcpitch-east)    -- node {\midarrow} (Fcpitch-east -| cat1-south) -- node {\midarrow} (cat1-south);

\draw [connection]  (Fccount-east)    -- node {\midarrow} (Fccount-east -| cat1-south) -- node {\midarrow} (cat1-south);


\pic[shift={(0.75,0,0)}] at (cat1-east) 
    {RightBandedBox={
        name=Fc2,
        caption=,%
        xlabel={{"Fc+softmax","dummy"}},
        fill=\FcColor,
        bandfill=\FcReluColor,%
        height=5,
        width=1,
        depth=1
        }
    };

\pic[shift={(0,0,0)}] at (Fc2-east) 
    {Box={
        name=softmax,%
        caption=,
        opacity=0.8,
        fill=\SoftmaxColor,%
        height=5,
        width=1,
        depth=1
        }
    };

\draw [connection]  (cat1-east)    -- node {\midarrow} (Fc2-west);

\end{tikzpicture}
\vspace{-30pt}
\caption{Deep neural network architecture for predicting at-bat outcomes. Input tensors are in blue, conv layers in yellow, max-pooling in red, and fully-connected layers in purple.  The circle concatenates the fully connected ``embedding'' layers.  Final softmax layer is in dark purple.  All convolutional layer activations are ReLU, while activations in Fc (fully-connected) layers are sigmoid.}
\label{F:outcome_pred}
\end{figure}
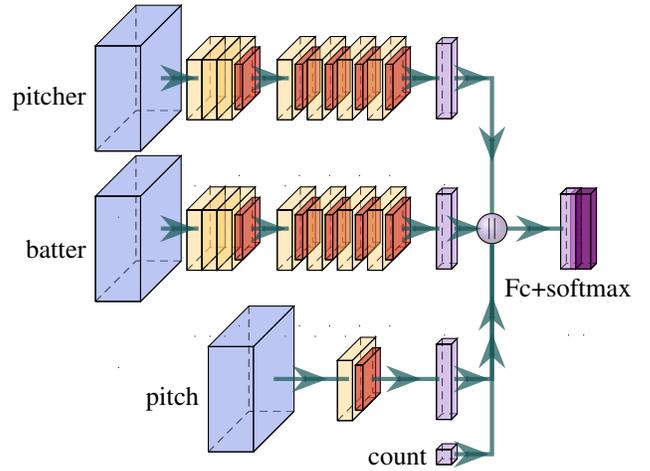

Next, we propose a novel deep convolutional neural network architecture for predicting outcomes shown in Figure~\ref{F:outcome_pred}.
While the primary innovation is in our representation of batter and pitcher as inputs which we described above, it is not a priori self-evident how to combine them into a single neural network architecture.
This problem is reminiscent of sensor fusion~\citep{Feng19}, where the architecture is often intricate, and the ideas do not necessarily transfer here.
What we opt for is essentially a \emph{late fusion} architecture.
Specifically, as illustrated in Figure~\ref{F:outcome_pred}, we first create separate convolutional architectures separately for the pitcher,  batter, and pitch, which (after convolutional layers) ultimately embed each into real vectors.
These are then concatenated with the pitch vector $z$ and the count representation $s$, with the resulting vector passed through several fully connected layers and, eventually, through the softmax layer to obtain the probability distribution over the four outcomes.

\subsection{Learning Pitcher Control}

A typical dataset of baseball at-bats consists of pitches thrown and documentation about where they ended up upon crossing the home plate, whether the batter swung, and the outcome.
What is conspicuously missing from such data is \emph{the intended location of the pitch}.
Given this state of affairs, it appears at first hopeless to get any handle on the distribution $D_l$ of actual pitch locations, given expected locations.
However, we now leverage some baseball conventional wisdom---imperfect, to be sure, but generally quite reasonable---that in certain counts the pitcher just wants to throw a strike.
In particular, suppose that we have a 3-0 count.
If the pitcher fails to throw a strike, and the batter does not swing, the batter is automatically on base.
It is, therefore, generally expected that a pitcher intends to throw a strike.
Now, \emph{where} in the strike zone the strike would be thrown remains uncertain, but here we add another reasonable assumption: if the pitcher aims to throw a strike, they will typically aim at the same part of the zone.
The final crucial assumption we make is that the spatial distribution of the \emph{error} (in 2 vertical dimensions of the strike zone) is zero-mean Gaussian, and only depends on the pitch type (but not, for example, count).

With the assumptions above, if we aggregate pitches thrown on 3-0 count for a particular pitcher, we can estimate a mean and the co-variance matrix Gaussian distribution, which in this case consists of 5 parameters for each pitch type $p$: $(\mu_x^p,\mu_y^p, \sigma_x^p, \sigma_y^p, \sigma_{xy}^p)_{p \in P}$, where $x$ is the horizontal and $y$ the vertical dimension of the strike zone viewed from the pitcher's perspective, with the origin in its center.

The challenge with the approach above is that we do  not necessarily have sufficient data for all pitchers on 3-0 counts \emph{for every pitch type} to effectively estimate the Gaussian distribution.
To address this issue, we use a deep neural network regression to estimate the Gaussian error model parameters for each pitcher and pitch type.
For this, we use the same pitcher representation as described in Section~\ref{S:predoutcomes} as input.

\subsection{Representing and Learning Batter Patience}

We represent the problem of batter ``patience'' as the task of learning the probability that the batter swings on a pitch in a particular area outside of the zone.
For this purpose, we first split our zones outside the strike zone further, with the presumption that when the pitch is far outside the zone, most batters will not swing, and the variation among batters stems primarily when pitches are relatively close to the strike zone.
For each such \emph{borderline} location $l \in Z_b$ and pitch $p \in P$, we learn a binary classifier to predict whether a given batter $x_b$, represented exactly as described above, will swing.
We use the same neural network architecture as the ``batter'' portion of the network in Figure~\ref{F:outcome_pred} for this purpose, and learn a separate neural network for each pitch and location.

\subsection{Solving the At-Bat Game}

\citet{Littman94} presented a general approach for solving finite zero-sum stochastic games which blends value iteration with linear programming.
We provide it here for completeness.
Note that in general, an equilibrium in a stochastic game entails randomization (\emph{mixed strategies}), and a mixed-strategy equilibrium in which policies depend only on current state, always exists~\cite{Filar12}.
For our purposes, it suffices to introduce notation for the pitcher's mixed strategies.
Let $\Delta(A_p)$ denote the set of mixed strategies for the pitcher.
We denote a particular mixed strategy (probability distribution) by $r_p \in \Delta(A_p)$.

The value function $V_i(s)$, which is initialized arbitrarily at iteration $i=0$, is updated as follows in iteration $i>0$:
\begin{equation}
\label{E:zsgame}
V_i(s) = \min_{r_p \in \Delta(A_p)}\max_{a_b \in A_b} \sum_{a_p \in A_p} r_p(a_p) U_i(s,a_p,a_b),
\end{equation}
where $r_p(a_p)$ denotes the probability of choosing $a_p$ under the distribution $r_p$ and
\[
U_i(s,a_p,a_b) = \left[ u(s,a_p,a_b) + \sum_{s' \in S} T_{ss'}^{a_p,a_b} V_{i-1}(s') \right].
\]
Now, we can observe that in any iteration $i$, $U_i(s,a_p,a_b)$ is fixed for every state, and therefore Equation~\eqref{E:zsgame} defines a zero-sum game, which can be solved for the equilibrium mixed strategy for the pitcher, $r_p$, using linear programming~\cite{Shoham08}.
\section{Experiments}

\paragraph{Experiment Setup}

To train and test our models, we use a Kaggle dataset comprised of at-bats from 2015-2018 MLB seasons~\citep{Kaggle18}.
For each at-bat, the dataset includes the pitcher and batter, the pitch thrown, along with its final coordinates as it crosses the home plate, count, and outcome (hit, foul, ball, etc).
Our total dataset includes 2696132 pitches thrown over 730585 at-bats.
Our training/test splits for each model ensured no overlap in pitchers or batters between the two.
Experiments were run on a Macbook Air 2020 (1.1 GHz dual-core i3, 8 GB RAM) running macOS 10.15.5. Neural networks were built in TensorFlow 2.5.0, and all source code was written in Python.


Throughout, we evaluate our approaches using batches of at-bats involving pitchers and batters that we categorize as \emph{strong} (well above average), \emph{average} (around average), and \emph{weak} (well below average).
The categorization is based on ranking either pitchers or batters in terms of empirical OBP (for pitchers, this would be OBP against).
We present results for two such pairing groups: 1) strong pitchers vs.~weak batters and 2) weak pitchers vs.~strong batters.


\paragraph{Outcome Predictions}


We begin by evaluating the efficacy of our neural network model for predicting outcomes of thrown pitches conditional on the batter swinging. Our test set of these outcomes comprised 147799 pitches.
Since predictions are distributions, we compare the average predicted probabilities of the four outcomes (strike, foul, out, and hit) for two large batches of at-bats: one pairing a strong pitcher with a weak batter, and another pairing a weak pitcher and a strong batter.
Both the predicted probabilities and empirical observations, are averaged over all corresponding counts and at-bats from test data.
\begin{figure}[h!]
\centering
\begin{subfigure}{0.23\textwidth}
\includegraphics[width=\textwidth]{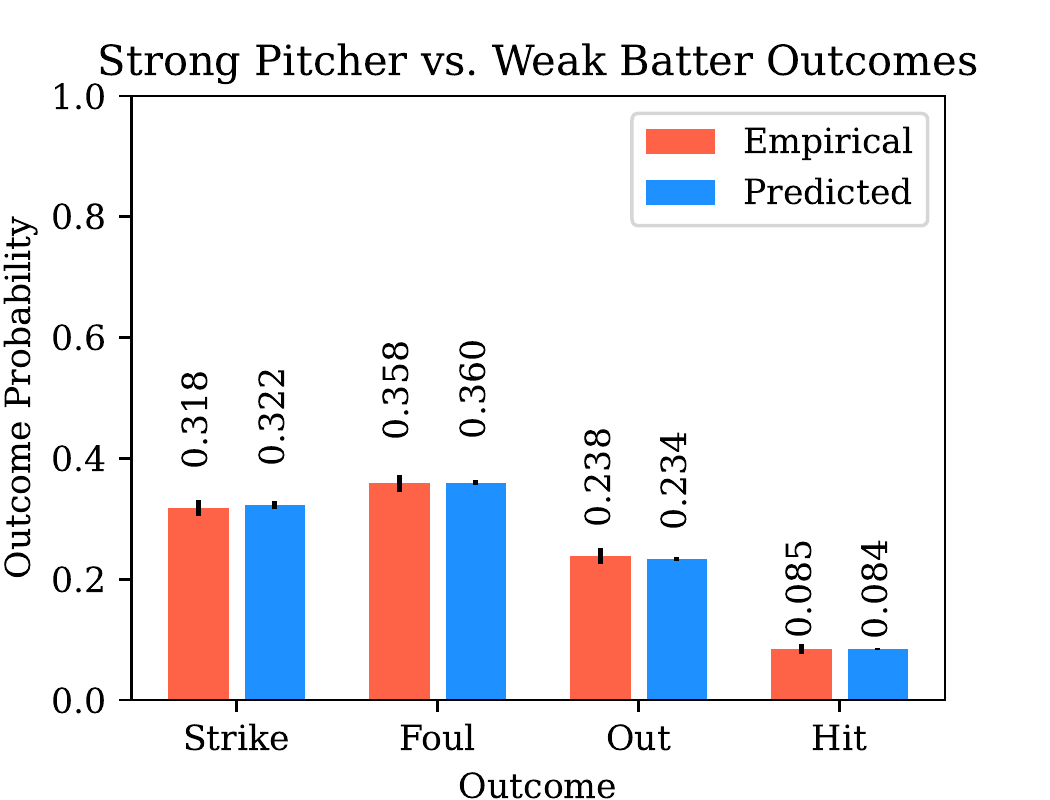}
\end{subfigure}
\begin{subfigure}{0.23\textwidth}
\includegraphics[width=\textwidth]{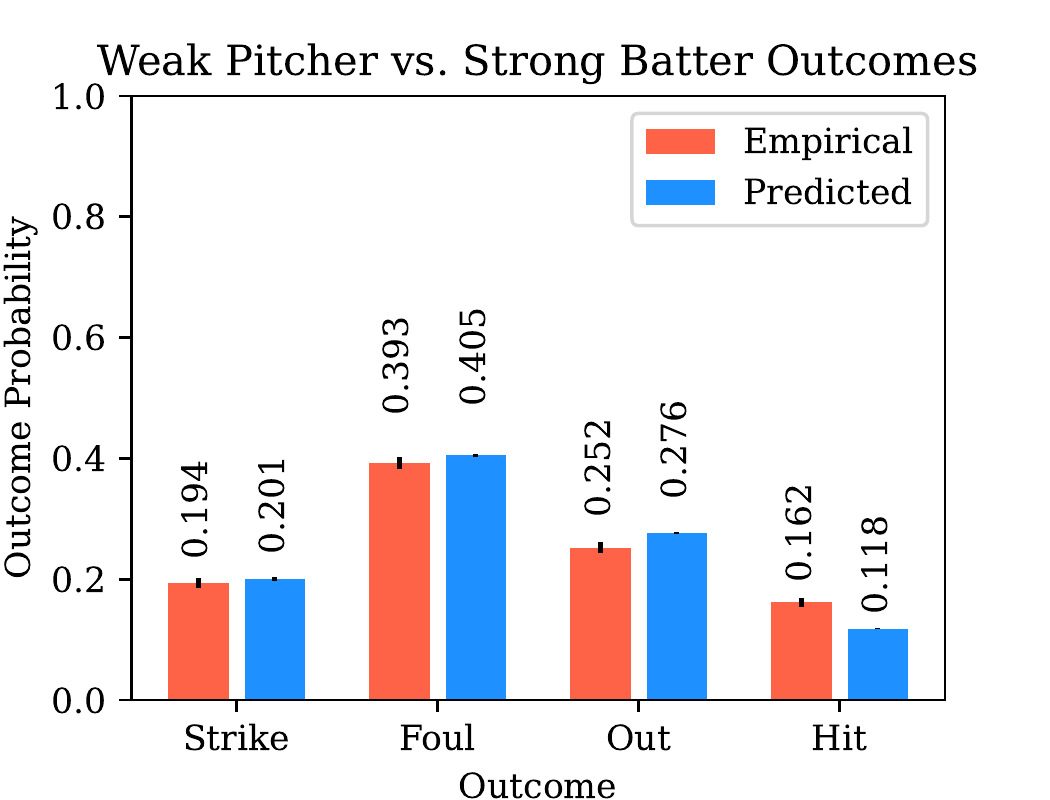}
\end{subfigure}
\caption{Average predicted probabilities and empirical observations of the four outcomes of swinging at a pitch (strike, foul, out, and hit).}
\label{F:outcome_matchups}
\end{figure}

The results are shown in Figure~\ref{F:outcome_matchups}.
First, we can observe that the predicted outcome distribution tracks empirical values very closely.
Of particular importance to us is that the relative probabilities are closely preserved: for example, foul balls are the most common whatever the matchup, and a strong pitcher facing a weak batter is far more likely to get a swing-and-miss strike than if the relative pitcher-batter strength is reversed.
\begin{figure}[h!]
\centering
\begin{subfigure}{0.23\textwidth}
\includegraphics[width=\textwidth]{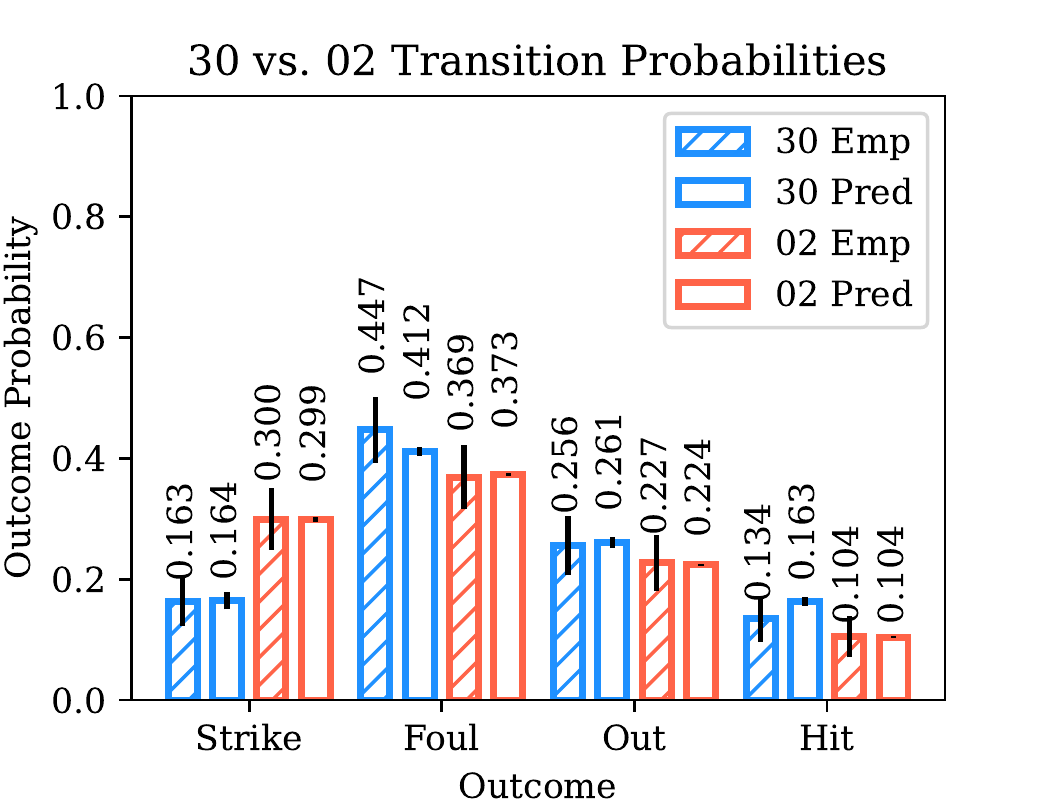}
\end{subfigure}
\begin{subfigure}{0.23\textwidth}
\includegraphics[width=\textwidth]{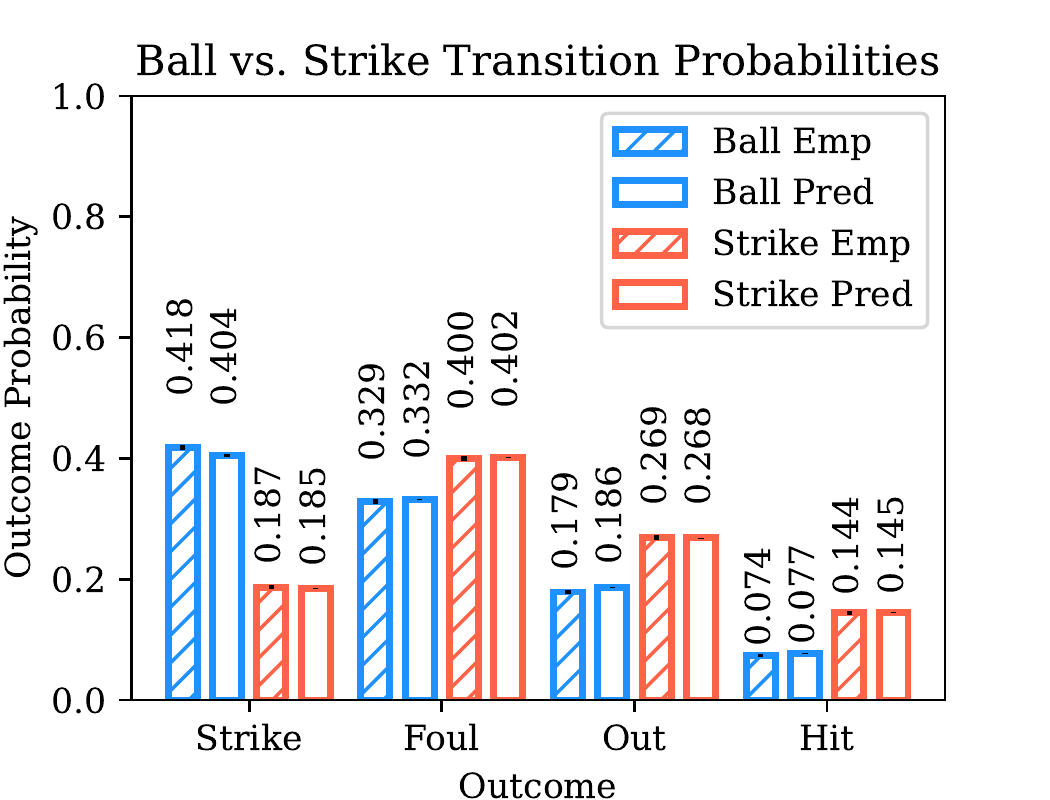}
\end{subfigure}
\caption{Average predicted probabilities and empirical observations of the four outcomes of swinging at a pitch (strike, foul, out, and hit).}
\label{F:outcome_count_location}
\end{figure}
Figure~\ref{F:outcome_count_location} offers additional demonstrations of the prediction efficacy in terms of tracking relative probabilities for different settings.
Comparing 3-0 and 0-2 counts, for example, one would expect a higher likelihood of a hit in the former than the latter, which we indeed observe (for both predicted and empirical distributions).
Similarly, we observe a significantly higher likelihood of a strike in an 0-2 than a 3-0 count, and we again observe this.
Analogously, contrasting outcomes when the pitch thrown ends up outside the zone (a ball) vs.~in the zone (a strike), we see a far greater likelihood of a swing-and-miss in the former case than the latter, but lower likelihood of a hit.

\paragraph{Control Predictions}

We evaluate our model of control in two ways: first quantitatively, and then qualitatively, using 118 pitchers observed in 3-0 counts.
First, we consider the accuracy of predicted Gaussian model covariance matrix parameters in terms of mean-squared error (MSE).
The MSE in the $x$-dimension is $0.036$, in the $y$-dimension it is $0.053$, and the MSE for the covariance of $x$ and $y$ is $0.025$.
These are quite small compared to empirically observed variances in either the $x$ (0.45-0.55) and $y$ (0.55-0.67) dimensions for the 4-seam fastball (these are considerably higher for other pitch types).
Second, we find that variance (both predicted and empirically measured, under the Gaussian assumption) is systematically higher for weaker pitchers.
This observation is congruent with baseball intuition, and suggests that the Gaussian model of pitcher error distribution is indeed a reasonable starting point.




\paragraph{Patience Predictions}
Next, we evaluate the efficacy of our approach of batter patience prediction over 126145 test pitches thrown outside of the strike zone, but sufficiently close for batters to swing at them.
\begin{figure}[h!]
\centering
\begin{subfigure}{0.23\textwidth}
\includegraphics[width=\textwidth]{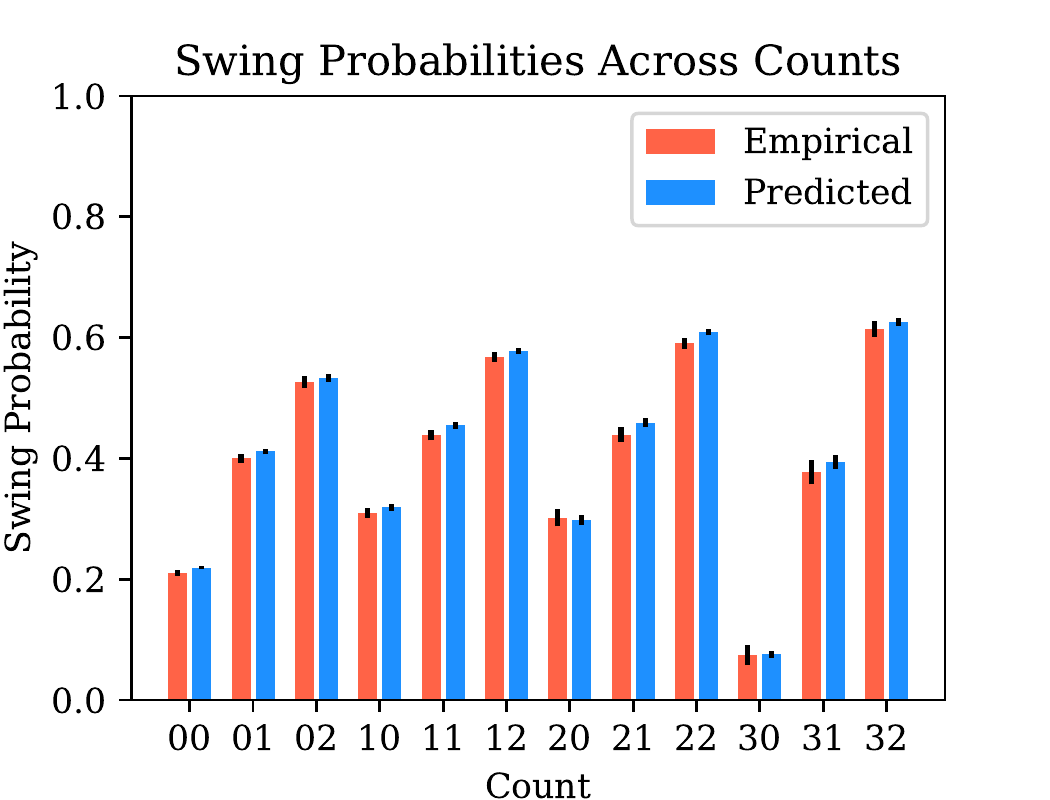}
\end{subfigure}
\begin{subfigure}{0.23\textwidth}
\includegraphics[width=\textwidth]{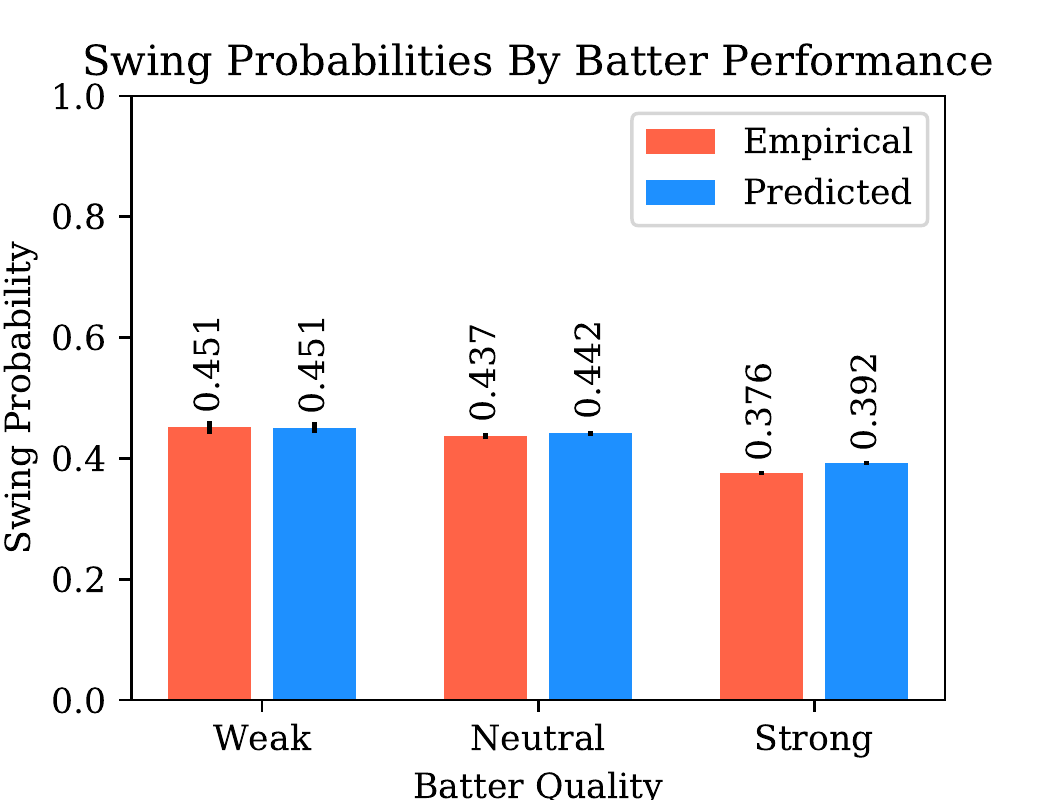}
\end{subfigure}
\caption{Average predicted probabilities and empirical observations of the four outcomes of swinging at a pitch (strike, foul, out, and hit).}
\label{F:patience}
\end{figure}
The results are shown in Figure~\ref{F:patience}, where we consider variation by counts (left), and by batters with differing degrees of success (right).
First, note that predictions are remarkably close to empirical swing probabilities.
Second, we can observe qualitative trends for both empirical and predicted swing propensities that are consistent with common baseball intuition.
For example, batters very rarely swing at pitches outside the zone on a 3-0 count, and very frequently on 2-strike counts.
Second, the top batters (those in our \emph{strong} category) are indeed noteworthy in that they tend to be more patient.



\paragraph{Overall Effectiveness}

Finally, we compare the effectiveness of the proposed \emph{stochastic game (SG)} approach in terms of on-based percentage (OBP) in equilibrium 
with empirical OBP.
Our first observation is that, overall, \emph{SG} significantly reduces OBP: over 3600 matchups on test data, the empirical OBP is 0.329, whereas the OBP in \emph{SG} equilibrium is \textbf{0.242}, more than a 25\% reduction!


\begin{figure}[h!]
\centering
\begin{subfigure}{0.23\textwidth}
\includegraphics[width=\textwidth]{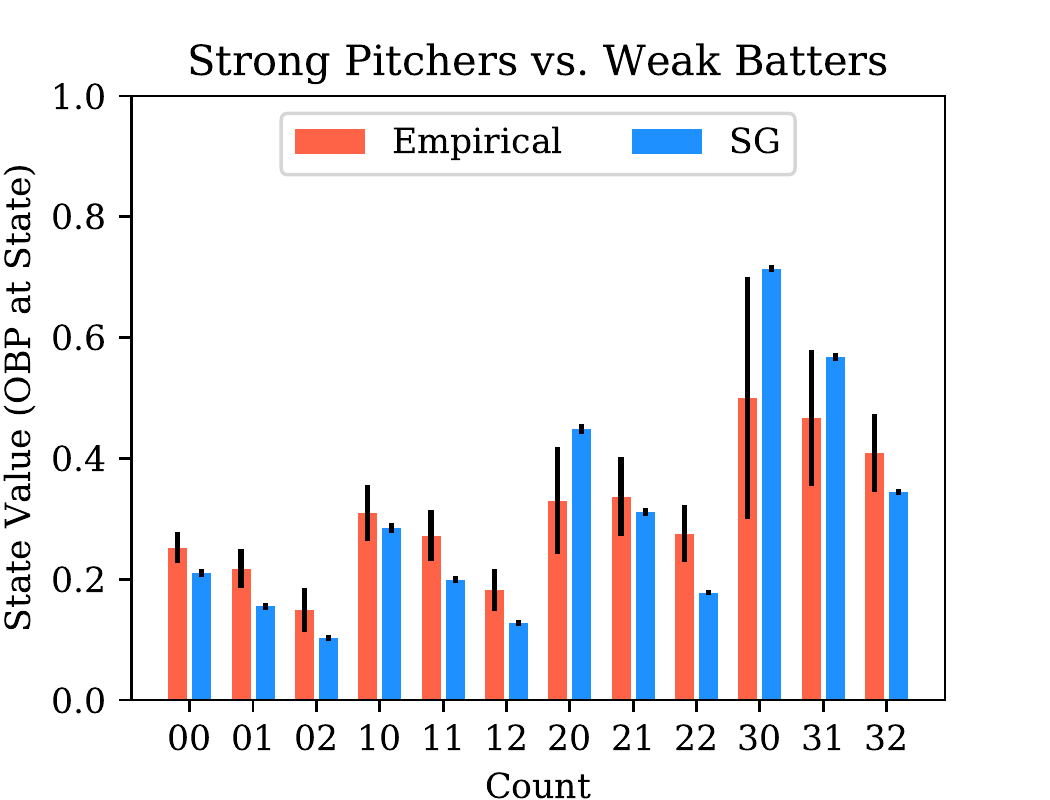}
\end{subfigure}
\begin{subfigure}{0.23\textwidth}
\includegraphics[width=\textwidth]{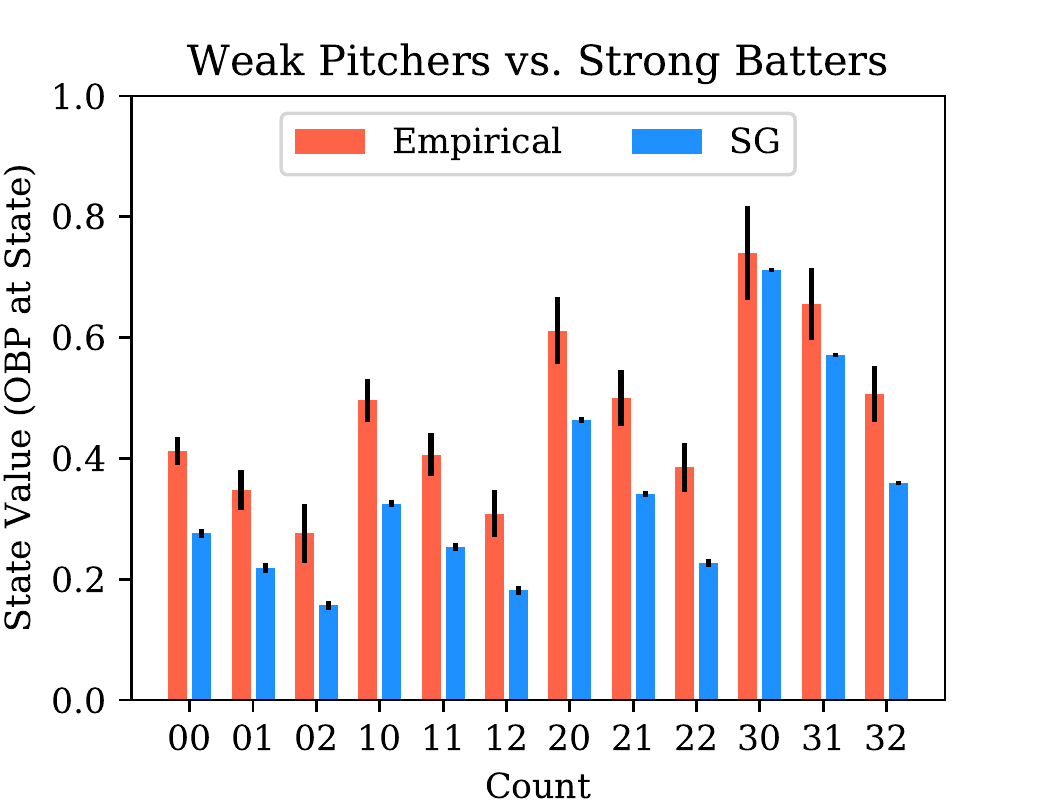}
\end{subfigure}
\caption{Stochastic game (SG) OBP vs.~empirical OBP, for different matchups and counts.}
\label{F:obp}
\end{figure}
In Figure~\ref{F:obp}, we delve deeper by comparing \emph{SG} and empirical OBP for different counts, and contrasting, in particular, two classes of matchups: 1) strong pitchers paired with weak batters, and 2) weak pitchers paired with strong batters.
First, note that we see expected variation for different counts: for both \emph{SG} and empirical OBP, counts that favor the batter (2-0, 3-0, and 3-1) have dramatically higher OBP than those that favor the pitcher (0-2, 1-2).
What is especially interesting, however, is that \emph{SG} does not exhibit much advantage in batter-favored counts.
However, \emph{SG} offers a considerably greater advantage when the count is roughly even, or the pitcher is ahead.
Furthermore, \emph{SG} holds a far greater advantage for weaker pitchers: for example, on 0-0 count, empirical OBP in these matchups is, on average, 0.412, while \emph{SG} yields an OBP of 0.276.
Indeed, we see this in a number of counts, with an advantage of \emph{SG} in a 1-0 and 1-1 counts particularly significant.




\section{Conclusion}

We propose a novel game-theoretic model of a baseball at-bat as a stochastic game.
The key challenge of this model is deriving the transition probability from empirical at-bat data for an arbitrary pitcher-batter pair.
We present a novel approach for doing so which is based on first decomposing the transition distribution into several parts, and propose deep neural network approaches for learning the key parts from data.
Our experiments demonstrate the efficacy of the proposed approach, showing, in particular, that the distribution of pitches and locations we compute in the resulting stochastic game yields significantly higher utility for the pitcher (lower on-based percentage for the batter) in nearly every count.
Moreover, we show that the proposed approach is particularly beneficial for average and below-average pitchers.

\bibliographystyle{aaai22}
\bibliography{main}

\end{document}


\maketitle
\section{Approach}

\paragraph{Pitch Tensor}
\label{S:predoutcomes}

\begin{figure}[h!]
    \centering
    \includegraphics[width=1.5in]{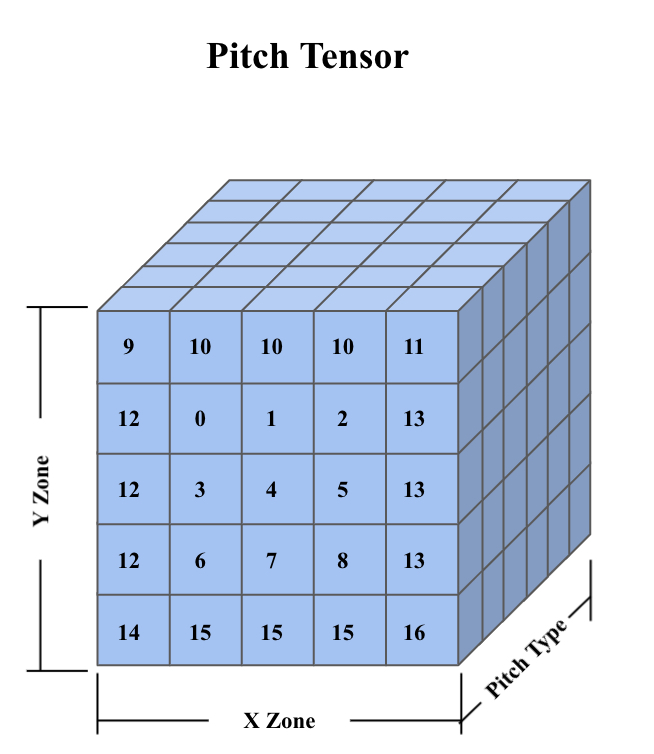}
    \caption{Pitch  tensor visualizations. Pitch tensors are of the shape (5,5,6) where the first two dimensions are used to physically represent the zone location. The final dimension represents pitch type.}
    \label{F:pitch_tensor}
\end{figure}

In our outcome and batter patience model inputs, we represent the pitch thrown as a three dimensional tensor. This tensor is comprised of all zero entries except for the index of the specific zone and pitch type combination, which is encoded as a one. This means any given pitch tensor will have a single one entry (or three one entries if the pitch is thrown to zones 10, 12, 13, or 15). See Figure~\ref{F:pitch_tensor} for a visual representation of a pitch tensor.

\paragraph{Pitch Distribution}

\begin{figure}[h!]
\centering
\begin{subfigure}{0.3\textwidth}
\includegraphics[width=\textwidth]{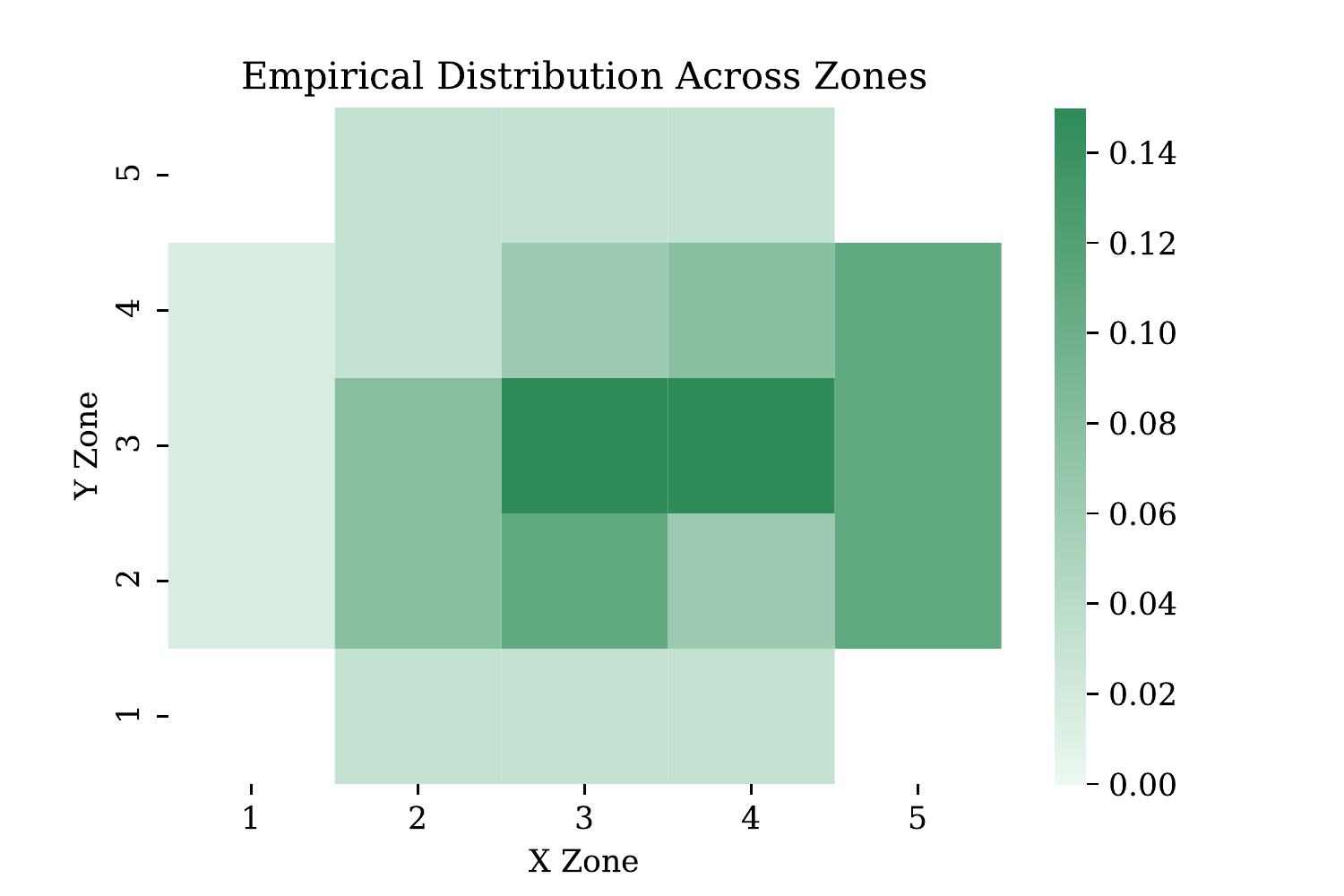}
\end{subfigure}
\begin{subfigure}{0.3\textwidth}
\includegraphics[width=\textwidth]{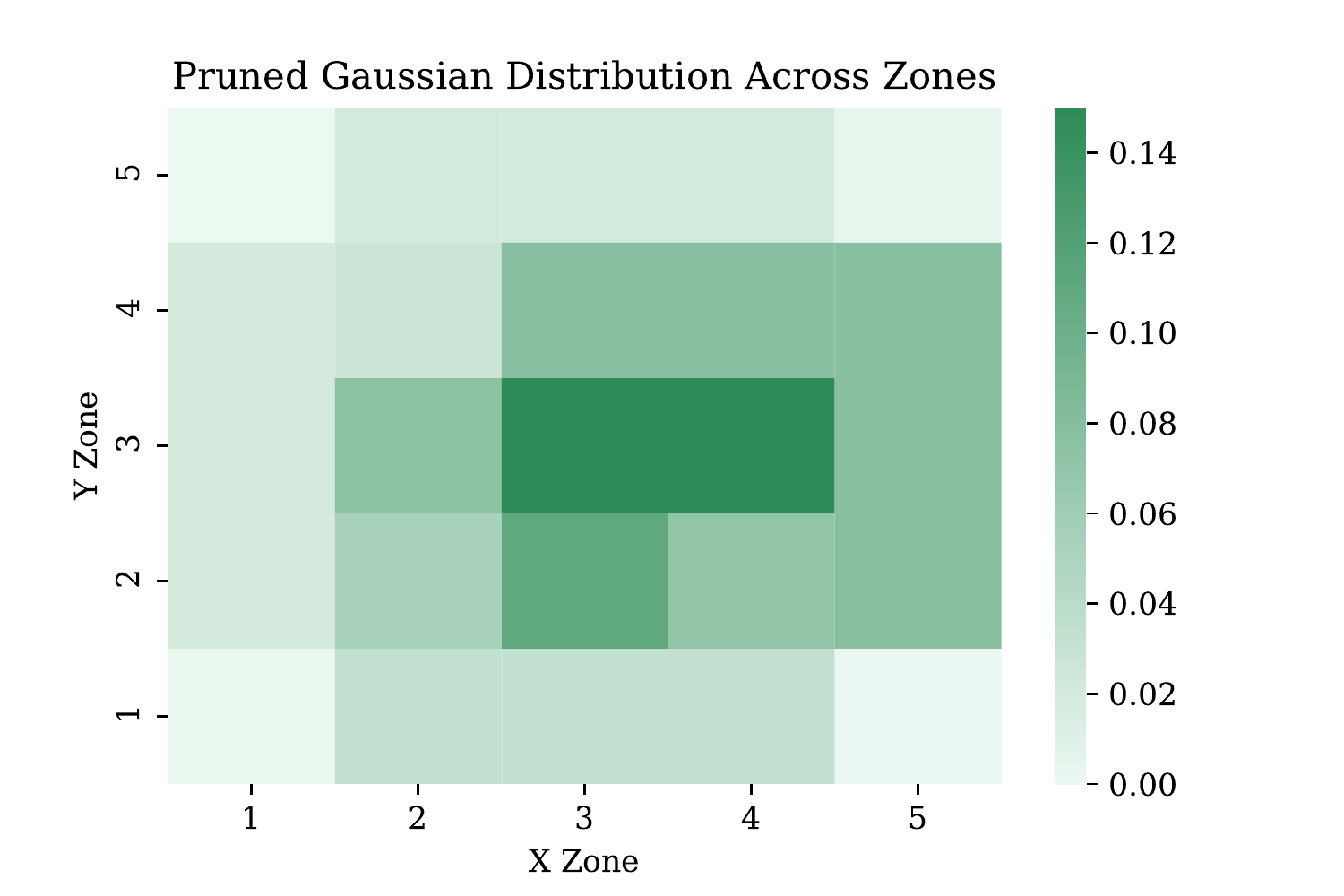}
\end{subfigure}
\caption{A heat map of pitch locations across zones for a fourseem fastball from an arbitrary pitcher, Geo Gonzalez. The top figure shows empirical pitches thrown at a 3-0 count. The bottom figure shows a sample (N = 100,000) from the pruned Gaussian learned from pitches thrown.}
\label{F:dist}
\end{figure}

In our pitcher control model, we assume that a pitcher's pitch distribution for a given pitch at a 3-0 count is Gaussian. We assume that this Gaussian is unimodal. First, we learn the maximum likelihood Gaussian for a given pitcher and pitch type at a 3-0 count (if \( n_{pitches} > 10\)). Then, we prune the least likely 5\% of these points (as determined by this original Gaussian) and regenerate the distribution on the remaining 95\% of points. In doing this, we attempt to systematically remove outlier pitches that do not represent the pitcher's true control (e.g., a pitcher throwing an outside ball to intentionally walk a batter). Figure~\ref{F:dist} compares samples from the pruned Gaussian learned from pitches thrown at a 3-0 count to the empirical distribution of these pitches for an arbitrary pitcher Geo Gonzalez. We see that these heat maps are, in fact, very similar, which signals that our assumptions about this distribution are reasonable.

\paragraph{Batter Patience}
We use a minimum swing probability to determine if a zone at a specific count for a given batter is \textit{obvious}. In an obvious zone, the optimal batter action is overridden to always \textit{take} the pitch ($a_b = \tau$). The value we choose for this threshold is .8. 

\begin{figure}[h!]
    \centering
    \includegraphics[width=3.25in]{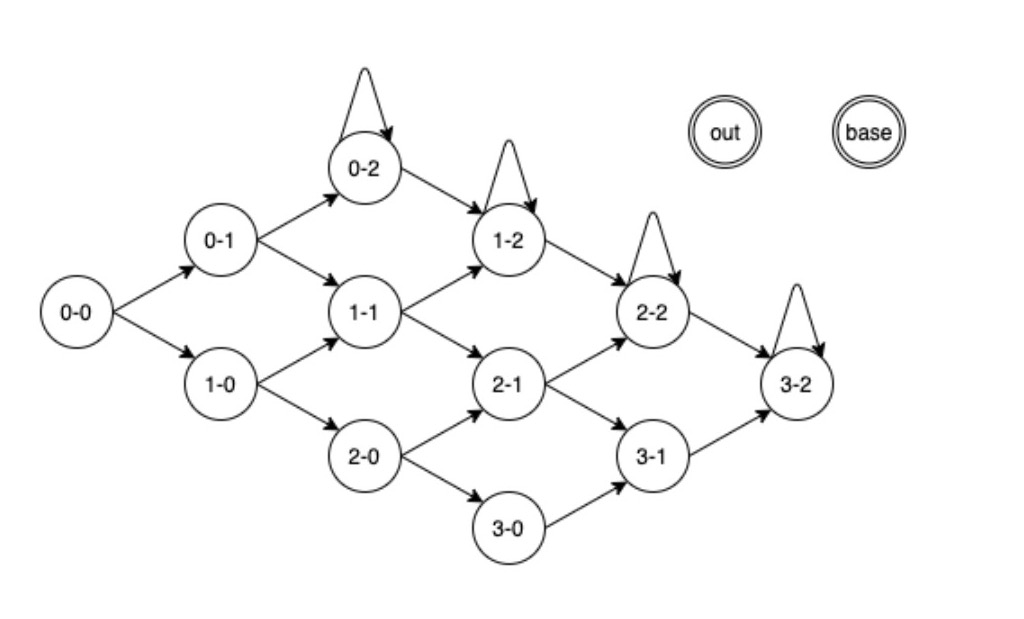}
    \caption{An illustration of counts (states) and how the game proceeds from one state to the next.}
    \label{F:counts}
\end{figure}

\paragraph{Stochastic Game States}
For a visual representation of the states, refer to Figure~\ref{F:counts}. This is a visual representation of how a game can progress from one count to another. A game can always progress to the terminal states of \textit{out} or \textit{on-base}. Note, when the strike count is at 2, a foul ball will not advance the state to any other count.

\section{Additional Experiments}

\paragraph{Control Predictions}

\begin{figure}[h!]
\centering
\begin{subfigure}{0.23\textwidth}
\includegraphics[width=\textwidth]{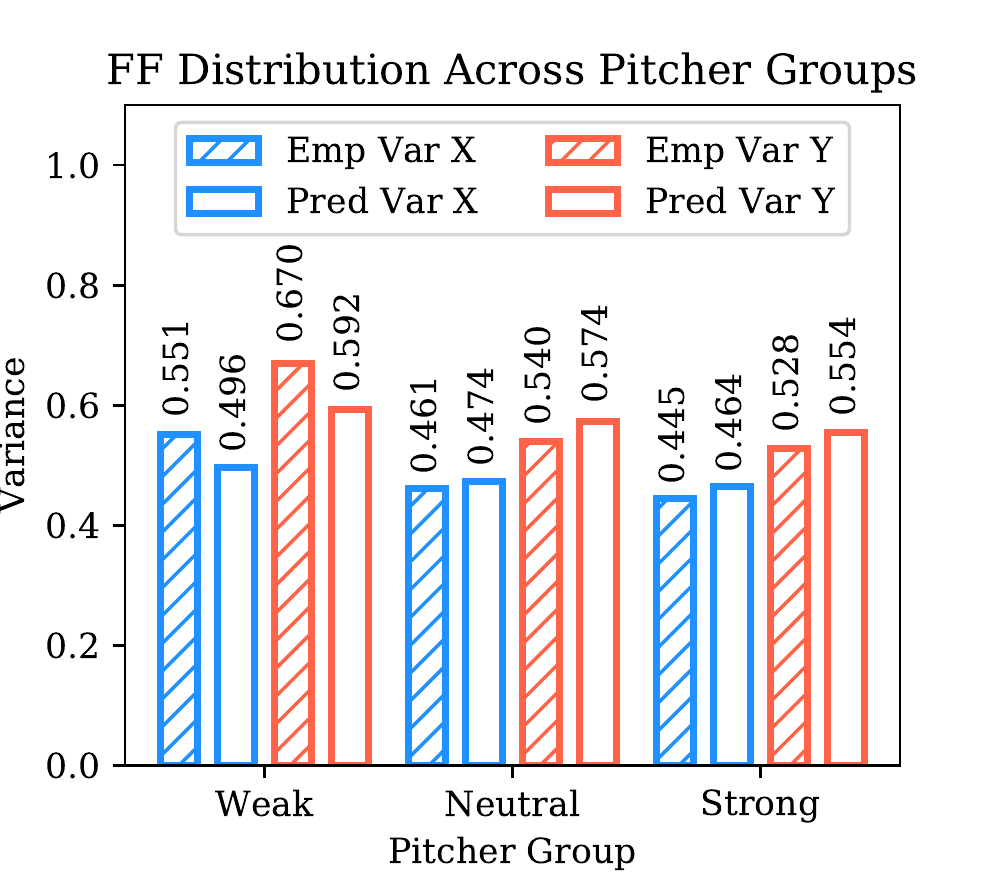}
\end{subfigure}
\caption{Pitch distributions by pitcher quality.}
\label{F:var}
\end{figure}

To demonstrate the ability of our control model to predict distributions, we include quantitative measures of distribution shape by pitcher quality. In our test set, we see that the higher the pitcher quality, the smaller the x and y variances of the predicted Gaussian are. Figure~\ref{F:var} highlights this trend. 

\begin{figure}[h!]
\centering
\begin{subfigure}{0.23\textwidth}
\includegraphics[width=\textwidth]{Strong_p_Weak_b_trans_probs.pdf}
\end{subfigure}
\begin{subfigure}{0.23\textwidth}
\includegraphics[width=\textwidth]{Weak_p_Strong_b_trans_probs.pdf}
\end{subfigure}
\begin{subfigure}{0.23\textwidth}
\includegraphics[width=\textwidth]{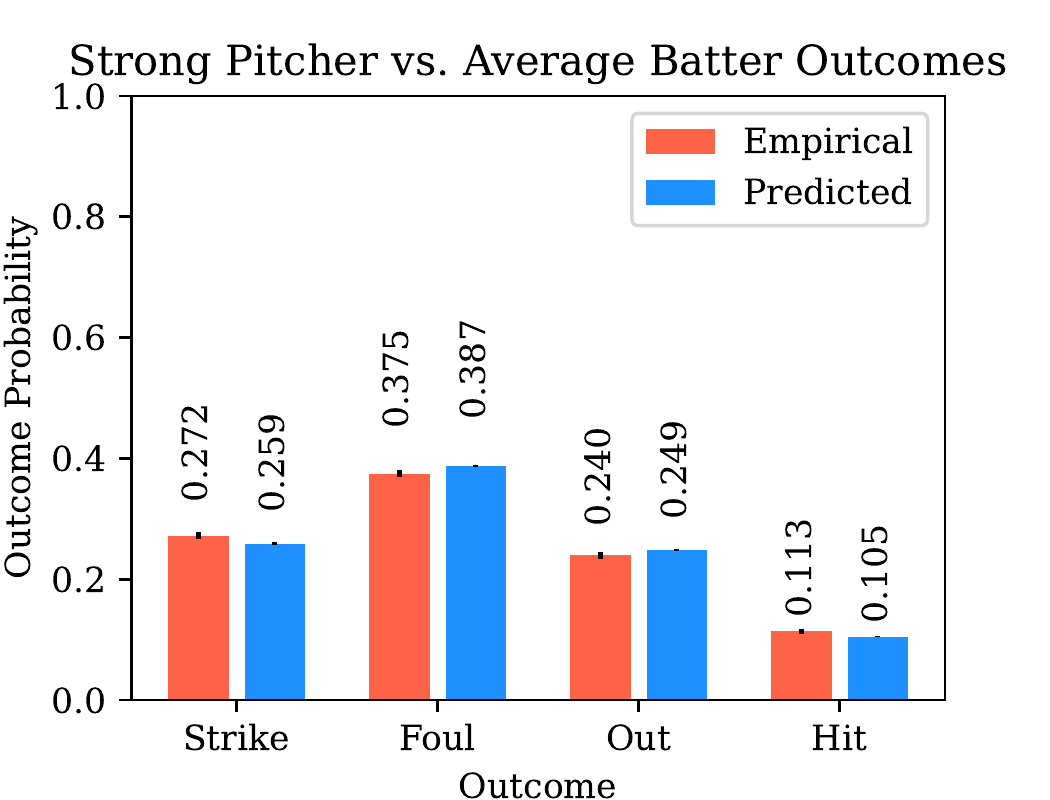}
\end{subfigure}
\begin{subfigure}{0.23\textwidth}
\includegraphics[width=\textwidth]{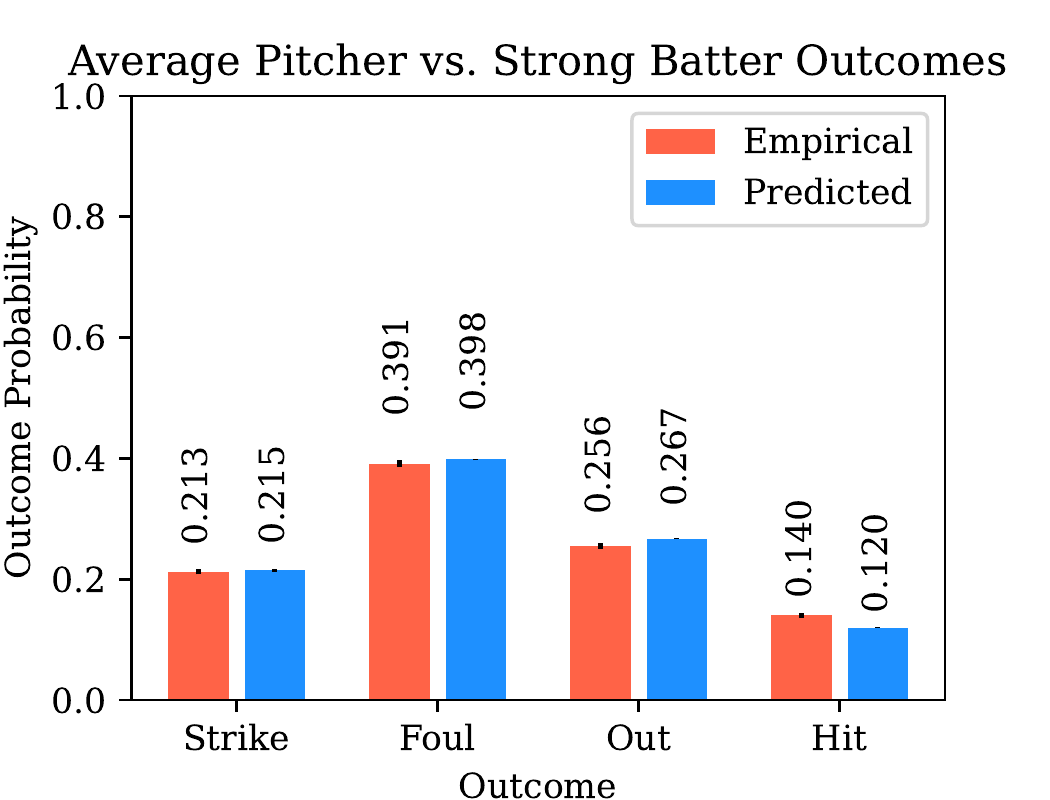}
\end{subfigure}
\begin{subfigure}{0.23\textwidth}
\includegraphics[width=\textwidth]{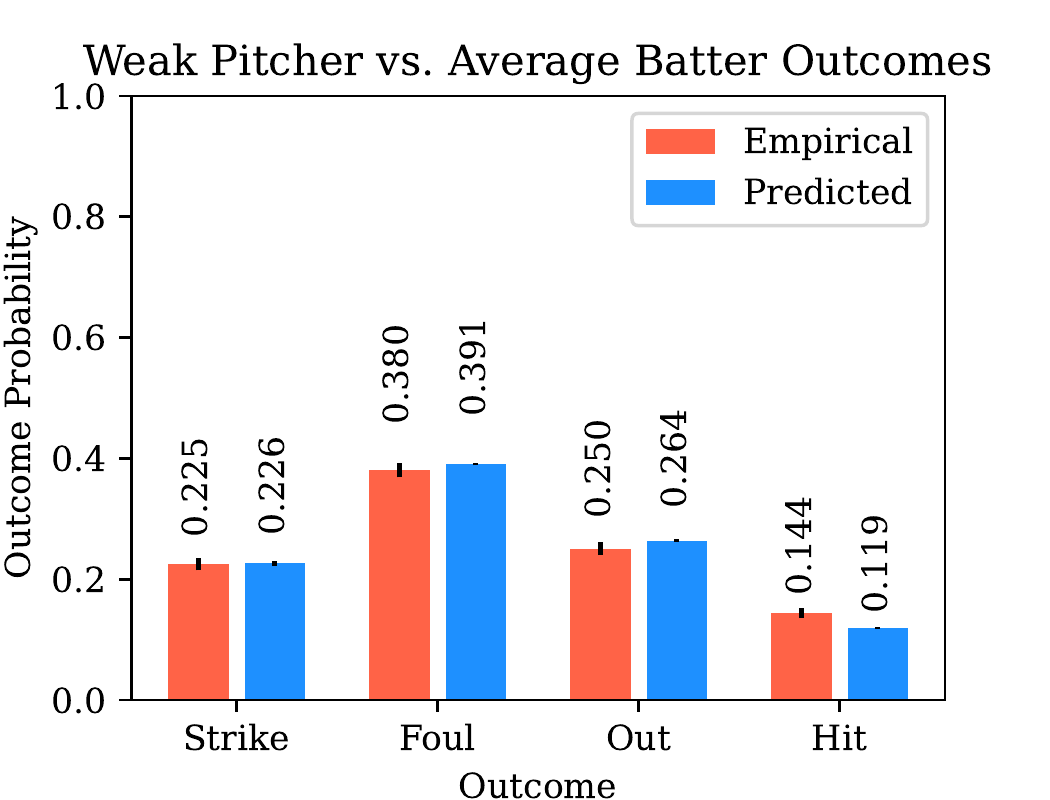}
\end{subfigure}
\begin{subfigure}{0.23\textwidth}
\includegraphics[width=\textwidth]{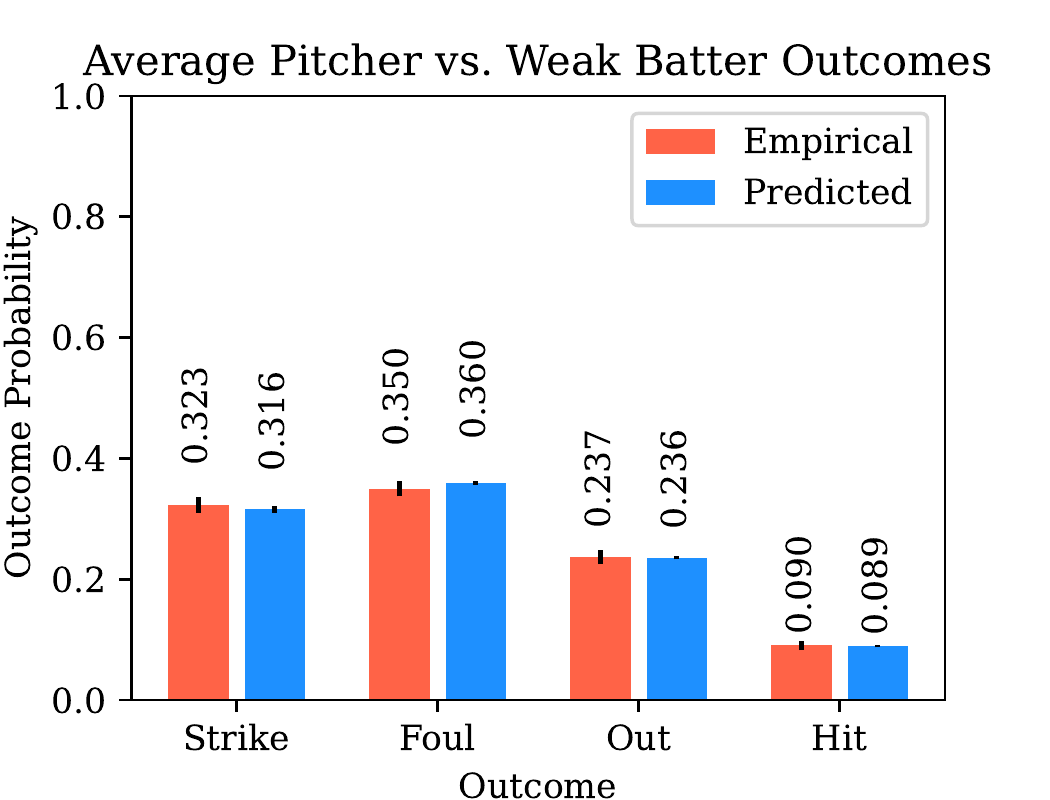}
\end{subfigure}
\begin{subfigure}{0.23\textwidth}
\includegraphics[width=\textwidth]{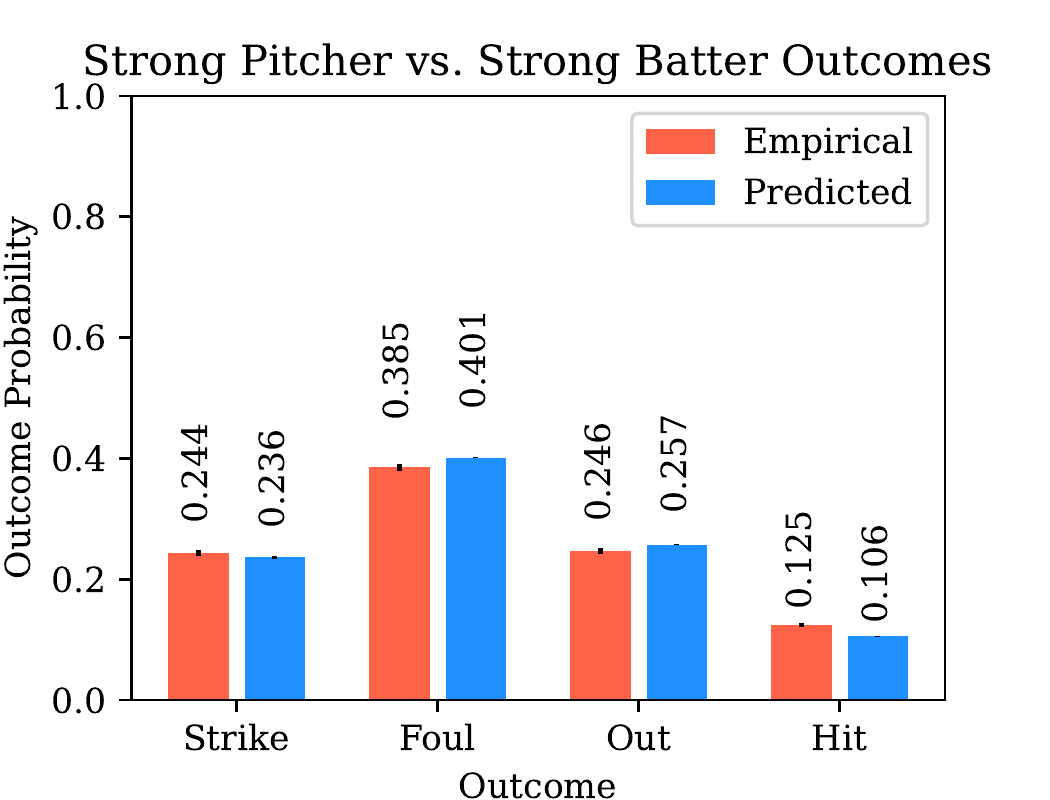}
\end{subfigure}
\begin{subfigure}{0.23\textwidth}
\includegraphics[width=\textwidth]{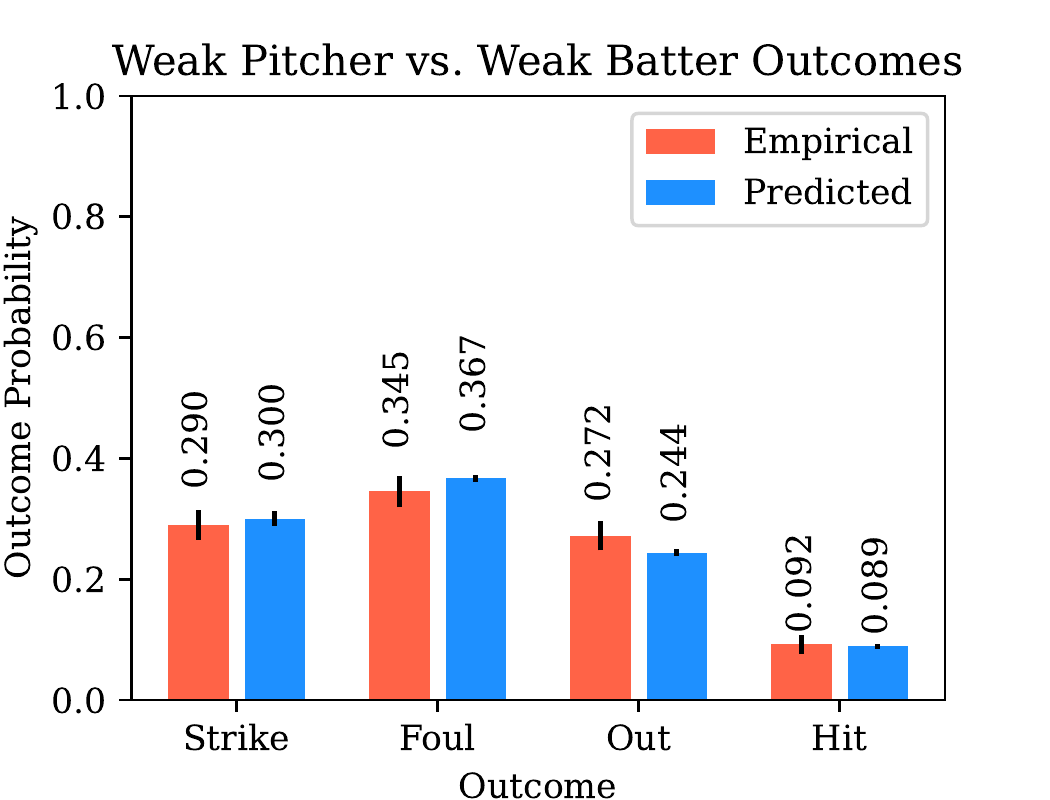}
\end{subfigure}
\begin{subfigure}{0.23\textwidth}
\includegraphics[width=\textwidth]{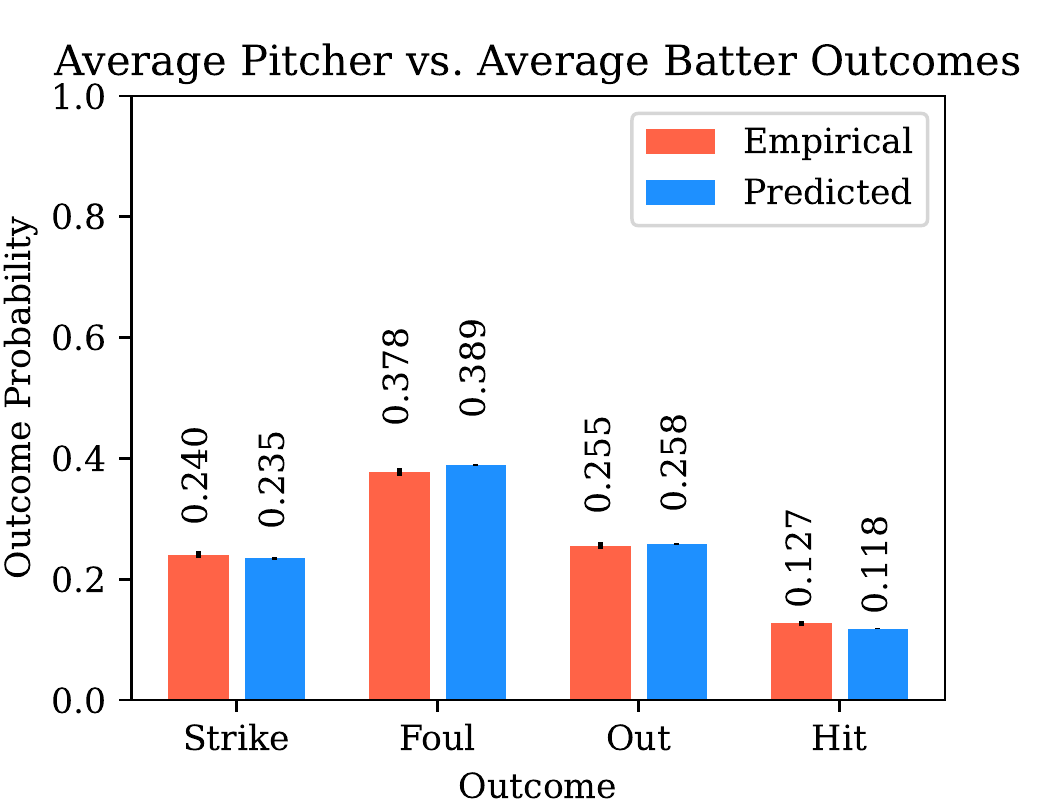}
\end{subfigure}
\caption{Swing outcome probabilities between all sample matchups grouped by pitcher and batter quality.}
\label{F:trans_prob}
\end{figure}

\paragraph{Outcome Predictions}
 In Figure~\ref{F:trans_prob}, we see the swing outcome probabilities across nine different grouping matchups. These highlight the ability of our model to effectively capture outcome probabilities based on the quality of both pitcher and batter. In stronger pitchers, we see relatively higher \textit{strike} probabilities, whereas in stronger batters, we see a greater probability of \textit{hit}. Across all matchups, the model's predictions closely match empirical outcome probabilities.

\begin{figure}[h!]
\centering
\begin{subfigure}{0.23\textwidth}
\includegraphics[width=\textwidth]{Strong_p_Weak_b_sg.pdf}
\end{subfigure}
\begin{subfigure}{0.23\textwidth}
\includegraphics[width=\textwidth]{Weak_p_Strong_b_sg.pdf}
\end{subfigure}

\begin{subfigure}{0.23\textwidth}
\includegraphics[width=\textwidth]{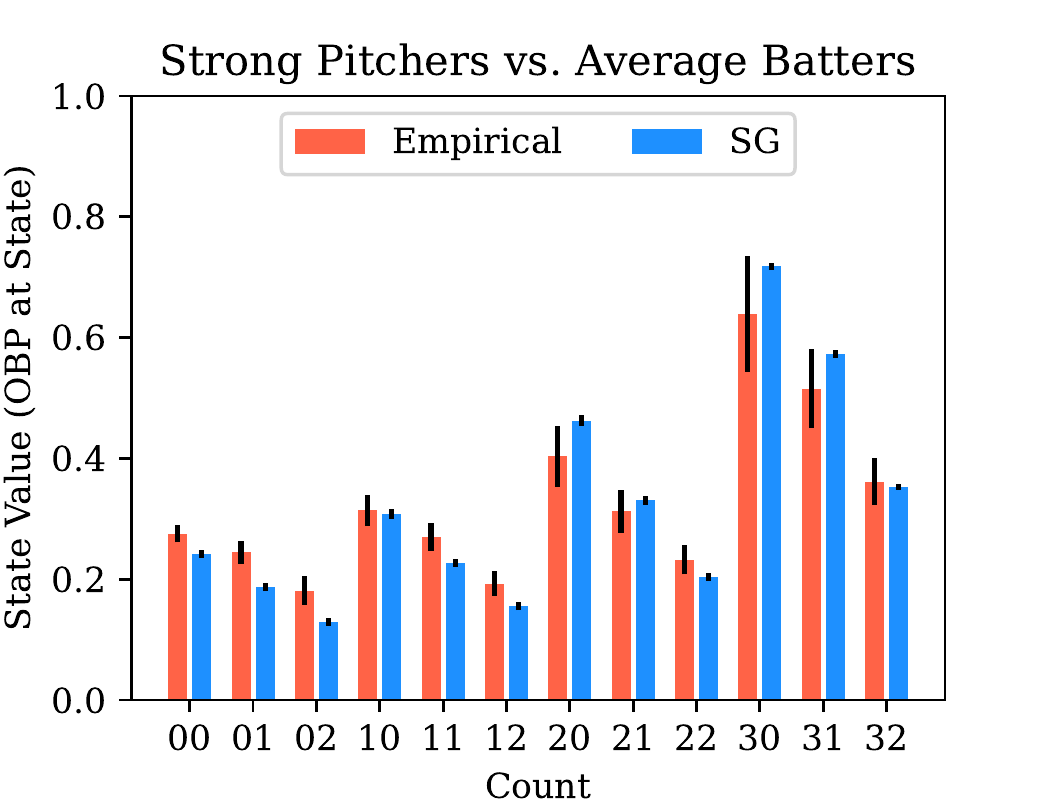}
\end{subfigure}
\begin{subfigure}{0.23\textwidth}
\includegraphics[width=\textwidth]{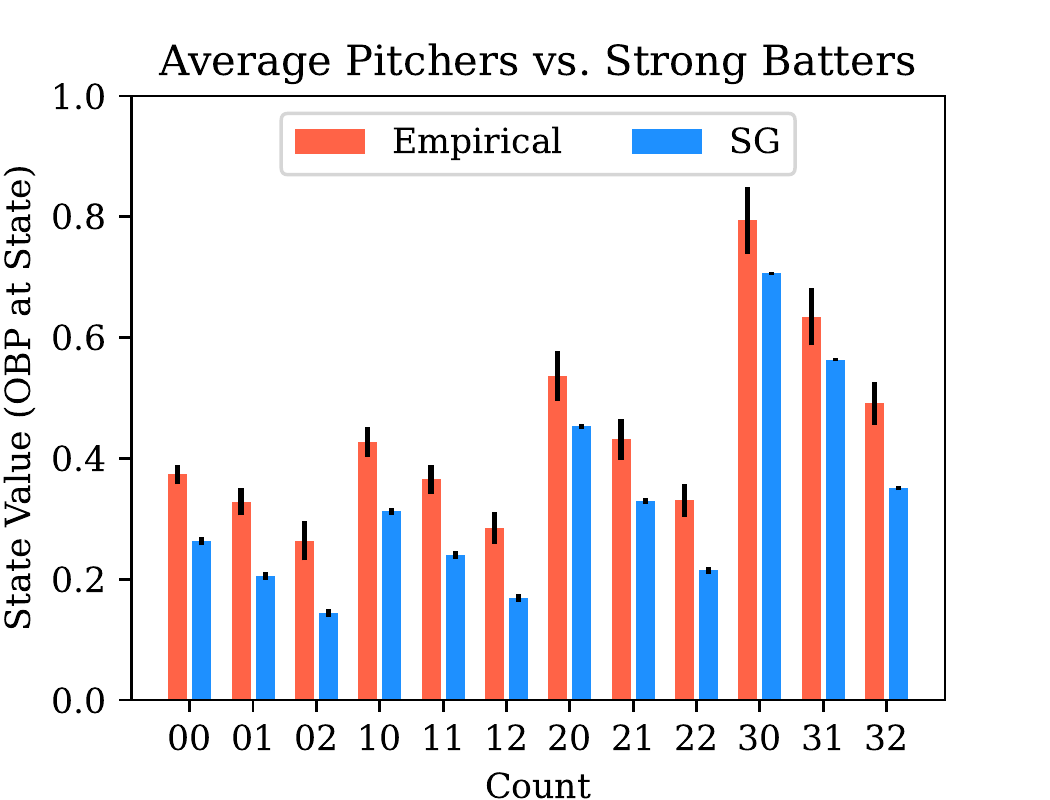}
\end{subfigure}
\begin{subfigure}{0.23\textwidth}
\includegraphics[width=\textwidth]{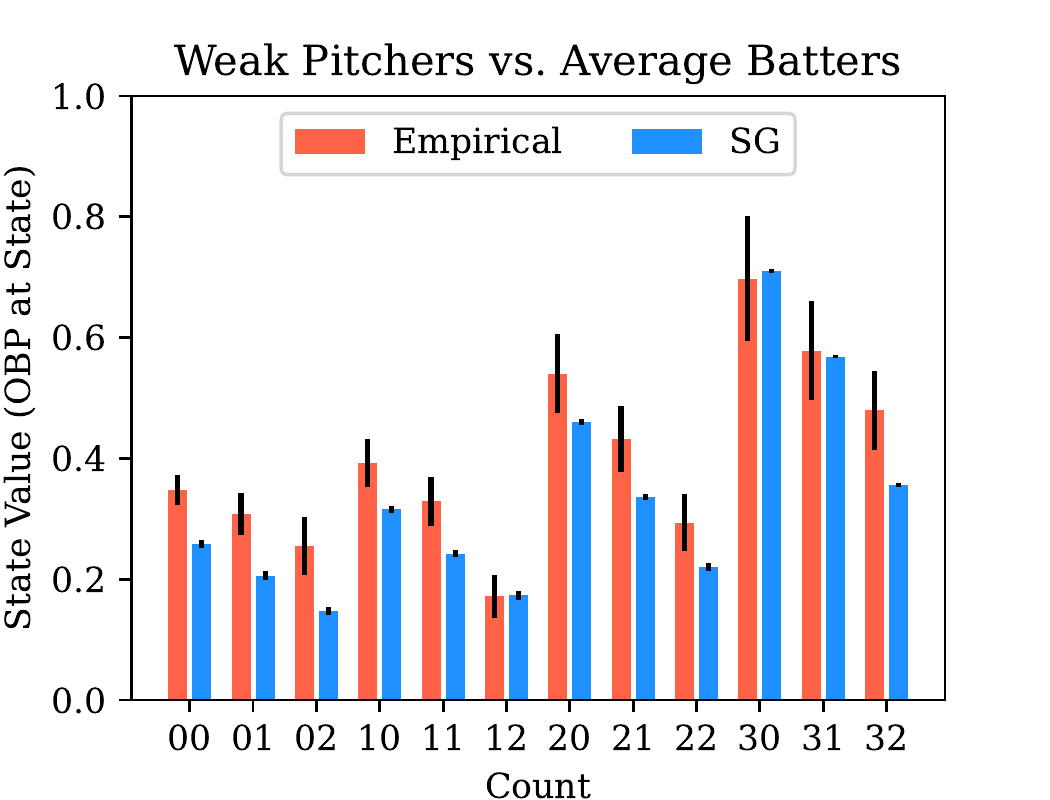}
\end{subfigure}
\begin{subfigure}{0.23\textwidth}
\includegraphics[width=\textwidth]{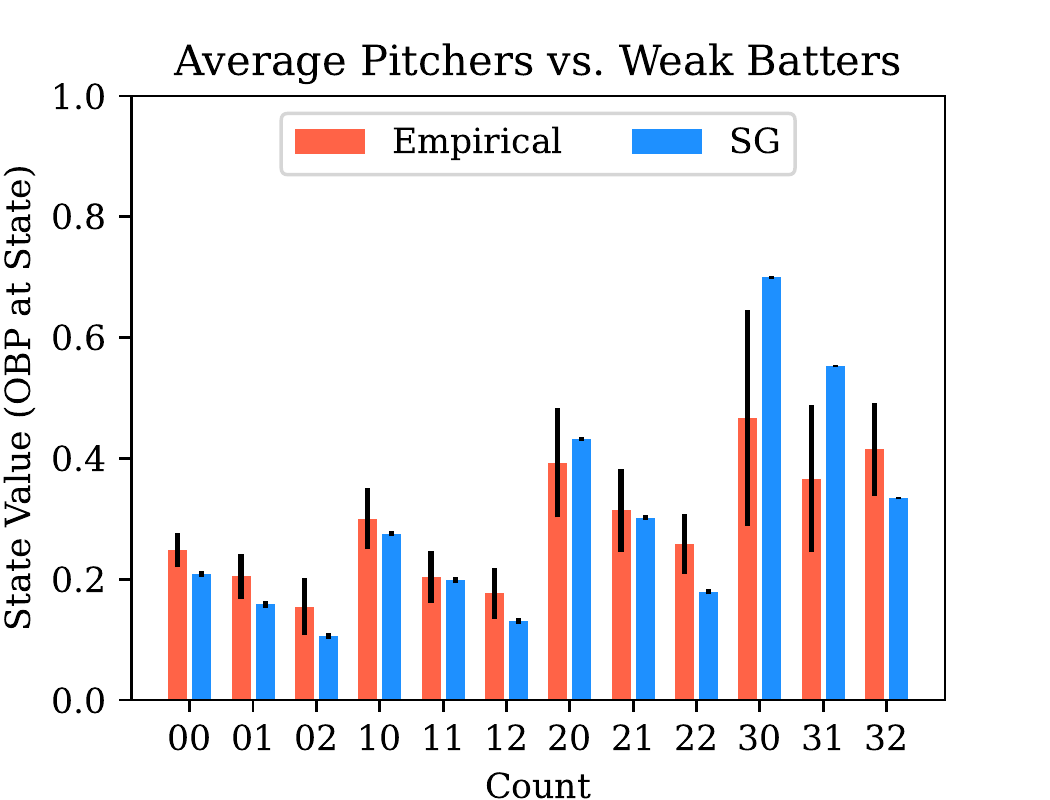}
\end{subfigure}
\begin{subfigure}{0.23\textwidth}
\includegraphics[width=\textwidth]{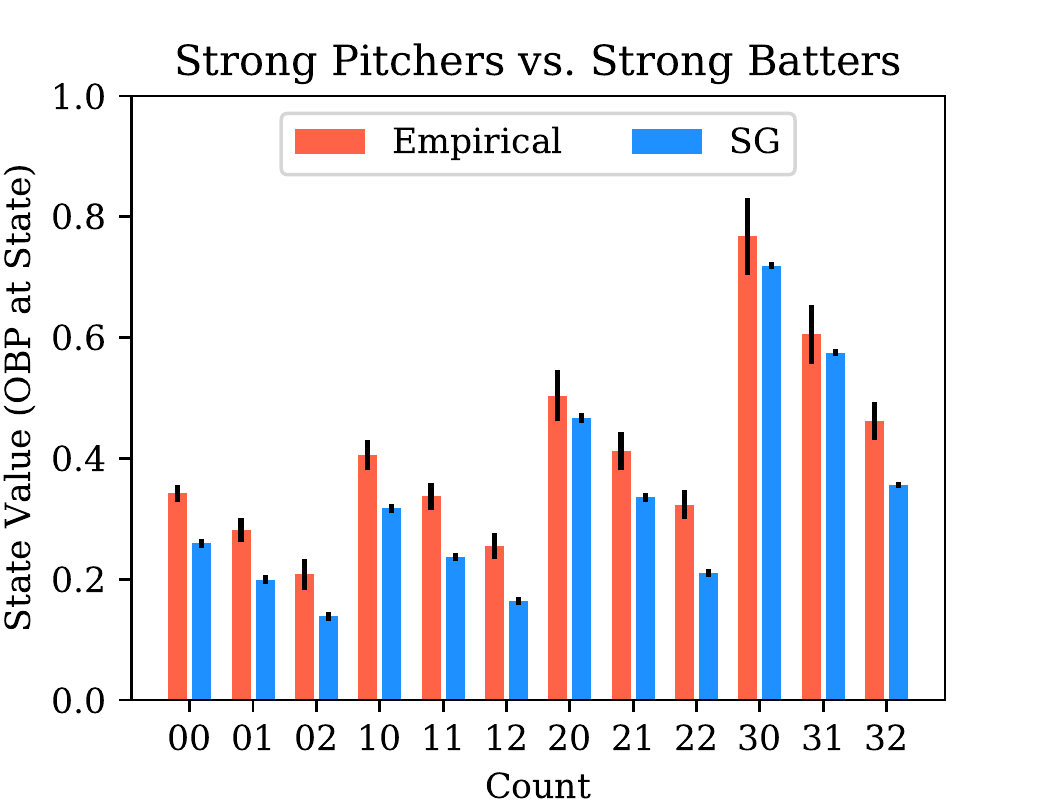}
\end{subfigure}
\begin{subfigure}{0.23\textwidth}
\includegraphics[width=\textwidth]{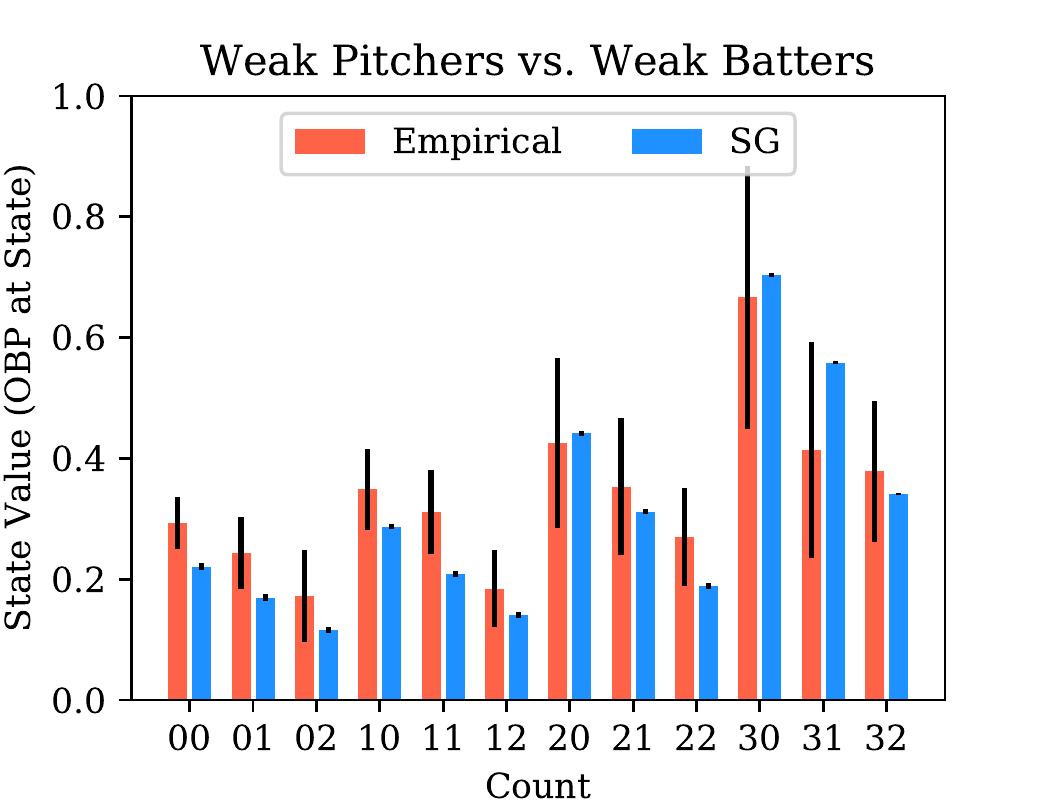}
\end{subfigure}
\begin{subfigure}{0.23\textwidth}
\includegraphics[width=\textwidth]{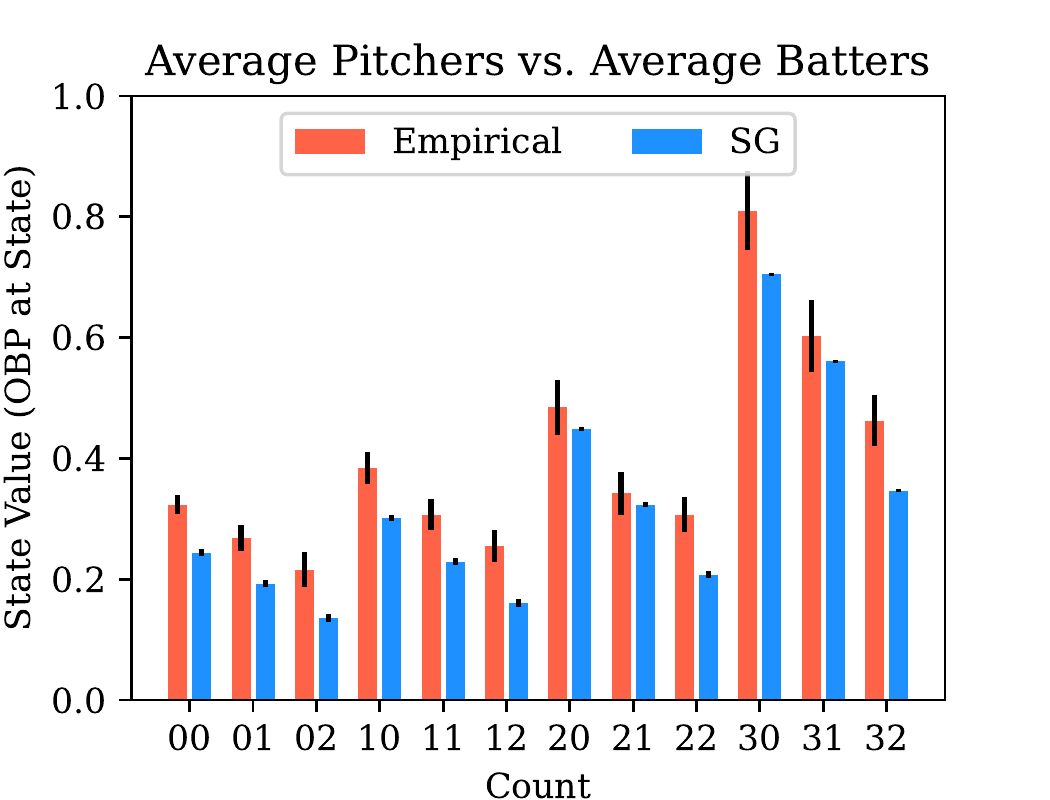}
\end{subfigure}
\caption{Stochastic game (SG) OBP vs.~empirical OBP, for nine different matchup combinations.}
\label{F:obp_counts}
\end{figure}

\paragraph{Overall Effectiveness}
In Figure~\ref{F:obp_counts}, we see the OBP values across counts for all nine grouping matchups. We note greater reductions in OBP at the 0-0 count (start of an at-bat) for relatively weaker pitchers. In addition, we notice wide confidence intervals empirically for strong pitchers in counts that favor the batter (especially 2-0, 3-0). This is because strong pitchers get into these counts less often, especially against weaker batters.

In Figure~\ref{F:SGoutputs}, we see an output of the stochastic game. This specific game is defined by the pitcher, Gerrit Cole, and batter, Chris Davis.

\begin{figure}[h!]
\renewcommand{\arraystretch}{1.5}
\begin{center}
\begin{tabular}{||c c r||}

    \hline
    Count & Value (OBP) & Policy\\
    \hline\hline
    00 & 0.2007 &
        \begin{tabular}{ c c }
         FF 4a & .341 \\ 
         SL 8a & .659  
        \end{tabular}
        
        \\
    \hline
    01 & 0.1367 &
        \begin{tabular}{ c c }
         FF 0a & .149 \\ 
         FF 1a & .700 \\
         CU 15a & .151
        \end{tabular}
        
        \\
    \hline
    02 & 0.0847 &
        \begin{tabular}{ c c }
         FF 0a & .373 \\ 
         SL 16a & .627 
        \end{tabular}
        
        \\
    \hline
    10 & 0.2697 &
        \begin{tabular}{ c c }
         FT 4a & .700 \\ 
         SL Fa & .300 
        \end{tabular}
        
        \\
    \hline        
    11 & 0.1818 &
        \begin{tabular}{ c c }
         FF 1a & .248 \\ 
         FF 4a & .700 \\ 
         FC 4a & .052
        \end{tabular}
        
        \\
    \hline
    12 & 0.1101 & 
        \begin{tabular}{ c c }
         FF 0a & .646 \\ 
         CU 15a & .354 
        \end{tabular}
        
        \\
    \hline 
    20 & 0.4410 &
        \begin{tabular}{ c c }
         FT 4a & .300 \\ 
         SL 4a & .700 
        \end{tabular}
        
        \\
    \hline 
    21 & 0.3025 &
        \begin{tabular}{ c c }
         FT 4a & .700 \\ 
         SL 4a & .300 
        \end{tabular}
        
        \\
    \hline
    22 & 0.1688 &
        \begin{tabular}{ c c }
         FF 1a & .652 \\ 
         CU 8a & .348 
        \end{tabular}
        
        \\
    \hline
    30 & 0.7112 &
        \begin{tabular}{ c c }
         FT 4a & .300 \\ 
         SL 4a & .700
        \end{tabular}
        
        \\
    \hline 
    31 & 0.5632 &
        \begin{tabular}{ c c }
         FT 4a & .300 \\ 
         SL 4a & .700
        \end{tabular}
        
        \\
    \hline 
    32 & 0.5632 &
        \begin{tabular}{ c c }
         FT 4a & .246 \\ 
         FC 4a & .054 \\
         SL 4a & .700
        \end{tabular}
        
        \\
    \hline 
\end{tabular}    
\end{center}
\caption{OBP value and policy output at each count from the stochastic game (SG) for a matchup of Gerrit Cole vs. Chris Davis.}
\label{F:SGoutputs}
\end{figure}